\documentclass[aip,reprint,pop]{revtex4-2}

\usepackage{graphicx}
\usepackage{subfigure}
\usepackage{epstopdf}
\usepackage{amsmath}
\usepackage{amssymb}
\usepackage{nicefrac}
\usepackage{csquotes}
\usepackage{dsfont}
\usepackage{dcolumn}
\usepackage{bm}
\usepackage[colorlinks=true, linkcolor=blue]{hyperref}

\begin{document}

\title{Wave drag in moving plasmas: recent developments and prospects}
\author{Renaud Gueroult}
\email{renaud.gueroult@laplace.univ-tlse.fr}
\affiliation{LAPLACE, Universit{\'e} de Toulouse, CNRS, INPT, UPS, 31062 Toulouse, France}
\author{Aymeric Braud}
\affiliation{LAPLACE, Universit{\'e} de Toulouse, CNRS, INPT, UPS, 31062 Toulouse, France}
\author{Julien Langlois}
\affiliation{LAPLACE, Universit{\'e} de Toulouse, CNRS, INPT, UPS, 31062 Toulouse, France}

\begin{abstract}
Wave propagation in a medium differs depending on whether this medium is at rest or moving with respect to an observer. Motion can notably lead to modifications of the wave trajectory, of its polarization, or of its transverse structure. Although these effects are well documented in isotropic dielectrics, they remain largely unexplored and unaccounted for in plasmas, despite the fact that simple models suggest they could in fact be large under certain conditions, as well as recent experimental observations. Here we first review existing models for motion effects on plasma waves, then identify a number of basic challenges that lie in the way of using these models to quantify motion effects in realistic configurations, and finally discuss possible workarounds.
\end{abstract}

\date{\today}
\maketitle

\section{Introduction}

Wave propagation in a medium differs depending on whether the medium is at rest or in motion with respect to the observer. The discovery of this fundamental effect goes back to the work of Arago, Fresnel, Stokes and Fizeau in the first half of the 19th century~\cite{Fresnel1818,Fizeau1851}. Although its origin was initially the source of confusion and debate, largely due to the erroneous assumption of a supporting aether, the effect of motion in moving isotropic dielectrics has since then been shown to be fully consistent with special relativity~\cite{vonLaue1907}. 

Beyond the original longitudinal drag~\cite{Fresnel1818,Fizeau1851} experienced by a wave propagating parallel to the direction of motion of a medium, subsequent studies of propagation in a moving isotropic dielectric showed that motion can also affect the wave trajectory (or ray) in the form of a transverse drag~\cite{Player1975,Jones1975}. Building on these findings but considering then a non-uniform motion, it was later shown that rotation is the source of a rotation of the wave polarization~\cite{Player1976,Jones1976} known as polarization drag~\cite{Fermi1923}, as well as of a rotation of the transverse structure of a wave known  as image rotation~\cite{Goette2007,Franke-Arnold2011}. Yet, despite these well documented effects of motion on wave propagation in isotropic dielectrics, the effect of motion on wave propagation in plasmas remains to date, other than for specific configurations mostly found in astrophysics~\cite{Lyutikov1998,Rafat2019}, largely neglected. This finding is all the more surprising given that plasma flows are ubiquitous, both in controlled laboratory experiments and in Nature.

Short of a definite explanation, a possible reason for this state of affairs is the scaling of motion effects in solid dielectrics. More specifically, drag effects in isotropic dielectrics typically scale as 
\begin{equation}
\label{Eq:scaling}
\beta\left(n_g'-\frac{1}{n'}\right),
\end{equation}
with $\beta=v/c$ the normalized velocity and $n'$ and $n_g'$ the refractive and group indexes in the medium's rest-frame. Because the velocity $v$ of moving dielectric is generally orders of magnitude smaller than the speed of light $c$, and because the wave and group index and of solid dielectrics are typically of order 1, drag effects are indeed small~\cite{Landau2013}. Quantitatively, as shown by Jones~\cite{Jones1975,Jones1976}, a solid glass rod rotating at a few hundred Hz will only deviate a light beam by a few tens of microradian after propagating a fraction of a meter. This is the reason why drag effects, despite having played an important role in the development of relativity~\cite{Whittaker1989}, have found little use in practical applications through most of the 20th century. 

Yet, while less immediately amenable to experiments, Eq.~\eqref{Eq:scaling} suggests that enhanced drag effects can in principle be obtained from greater $\beta$ or larger $n_g'$, which was later confirmed experimentally. Specifically, while fast motion is not a priori accessible in solid dielectrics due to mechanical constraints, much larger velocities can be achieved in dilute media. Following this route, it was for instance shown that the extreme angular frequency $\Omega\sim 10^{14}$~rad.s$^{-1}$ produced in molecular superrotors  can lead to a rotation of polarization of nearly one radian per meter ~\cite{Steinitz2020,Milner2021}. This is about $10^5$ times larger than that first observed by Jones in solid dielectrics~\cite{Jones1976}. Exploiting now the second option, and more particularly the artificially large group index $n_g'$ that can be obtained in stimulated media exhibiting slow-light conditions~\cite{Artoni2001,Franke-Arnold2011,Safari2016}, it was also shown that rotating a ruby rod at the same $100$~Hz angular frequency can then lead to a deviation of tens of degrees over a few centimeters~\cite{Franke-Arnold2011}, that is an increase by a factor of about $10^6$ compared to Jones' historical result~\cite{Jones1976}.  

This demonstration of enhanced drag effects, and the conditions for such enhancement, are interesting in the context of plasma waves for at least two reasons. First, plasmas support a wide range of waves, including some with low group velocity $v_g'=c/n_g'$. This is for instance the case of Alfv{\'e}n waves in subluminal conditions since in this case $v_g'=v_A\ll c$ with $v_A$ the Alfv{\'e}n speed. Second, plasma can feature fast flows, reaching velocities that are orders of magnitude larger than those achievable in solid dielectrics. This is clearly the case in astrophysics where relativistic flows are ubiquitous~\cite{Mirabel2003,Komissarov2012}, but fast flows can also be found in laboratory experiments: toroidal velocities of a few $100$~km.s$^{-1}$ are for instance observed in tokamaks, leading to angular frequencies $\Omega\sim 10^4-10^5$~rad.s$^{-1}$~\cite{Rice2007}. In light of literature in isotropic dielectrics, this combination of fast flows and large group indexes suggests that drag effects in plasmas could be significant~\cite{Gueroult2023}. Yet, Eq.~\eqref{Eq:scaling} on which this conjecture is made has only been derived for isotropic dielectrics, whereas plasmas are gyrotropic media. Quantifying the possible effects of motion on plasma waves thus requires a closer examination.

In this paper we review recent progress on the study wave propagation in moving media, with an eye to quantifying motion effects on waves in moving magnetized plasmas. In Section~\ref{Sec:Sec_II} we begin by recalling the main models that have been proposed to describe motion effects on plasma waves. In Section~\ref{Sec:Sec_III} we use more specifically Minkowski's theory to introduce what the main manifestations of motion on plasma waves are. With these results in hand we then identify in Section~\ref{Sec:Sec_IV} a number of issues that arise when modeling moving plasmas, and discuss what can be done to address them. Finally, Section~\ref{Sec:Sec_V} summarizes the main findings of this study.

\section{Models for waves in moving plasmas}
\label{Sec:Sec_II}

While, as mentioned in introduction, the effect of motion on wave dynamics appears to be neglected in all but a few plasma wave models used to date, it is important to note that the basic problem of the effect of motion on plasma wave has nevertheless already received some attention in the plasma literature. Although an exhaustive review of these contributions is beyond the scope of this study, we start by briefly discussing it here, both to provide some context and to introduce modeling frameworks that will be used later on. 

Formally, past studies on plasma waves in moving plasmas can be divided in three groups based on the approach employed. 

\subsection{Laboratory-frame perturbative method}

A first group consists of studies for which a wave equation is obtained in the laboratory frame $\Sigma$ considering a perturbation around a moving equilibrium as seen in $\Sigma$. This method is a direct extension of the standard linear perturbation theory used to model waves in static plasmas, and thus unsurprisingly was the first one to be considered in the literature~\cite{Bailey1948}. A downside of this method is that while it is well suited to describe waves in plasmas drifting in uniform $\mathbf{E}$ and $\mathbf{B}$ field since in this case the motion itself is not the source of currents, obtaining analytical solutions for the plasma linear response to a perturbation becomes much more intricate when there are currents associated with the equilibrium. This approach has nevertheless been applied to a number of plasma models of various complexity, including warm compressible isotropic plasmas~\cite{Unz1966}, cold two-fluid magnetized plasma~\cite{Unz1962}, gyrotropic compressible plasmas~\cite{Chawla1966} and Grad's 13-moment approximation of a warm two-fluid magnetized collisional plasma ~\cite{Talekar1974}.

\subsection{Rest-frame perturbative method}

The second category historically consists of studies that still consider the response to a perturbation, but this time in the plasma rest-frame $\Sigma'$. This technique was notably used to model waves in rotating plasmas, using momentum conservation equations which include inertial forces, and more particularly the Coriolis force following Chandrasekhar’s suggestion that it may play a predominant role in astrophysics~\cite{Chandrasekhar1953}. Here also a number of plasma models were used including MHD~\cite{Lehnert1954,Lehnert1955}, cold collisionless two-fluid~\cite{Tandon1966}, cold collisional two-fluid~\cite{Uberoi1970} and collisional compressible two-fluid~\cite{Verheest1974}. No matter the plasma model, an issue with this method is that one must decide on the form of Maxwell's equation in this non-inertial frame, leading to different results ~\cite{Engels1975}. Practically, another limitation of this method is that it gives access to observables as seen in the rest-frame $\Sigma'$, whereas observers are generally assumed to be at rest in the laboratory frame $\Sigma$.

\subsection{Minkowski's theory}

\begin{figure}
\begin{center}
\includegraphics[]{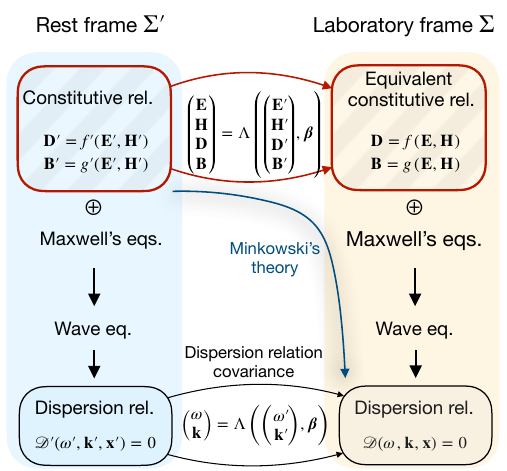}
\caption{Sketch of the transformations used in Minkowski's theory for waves in moving media. Equivalent constitutive relations in the laboratory frame $\Sigma$ are obtained by Lorentz-transforming the fields in the rest-frame constitutive relations, which are assumed to be known. These relations combined with Maxwell's equation provide a wave equation in $\Sigma$. Because of the complexity of the equivalent constitutive relations it is sometimes easier to obtain the laboratory frame dispersion relation $\mathcal{D}$ by directly Lorentz-transforming the wave 4-momentum $(\omega',\mathbf{k}')$ in the rest-frame dispersion relation $\mathcal{D}'$.}
\label{Fig:Minkowski}
\end{center}
\end{figure}

The third category consists of studies in which the wave dynamics is studied in the laboratory-frame $\Sigma$, though now through the use of equivalent constitutive relations obtained via a Lorentz-transformation of the electromagnetic fields in the rest-frame constitutive relations, as originally proposed by Minkowski~\cite{Minkowski1908}. This is illustrated in Fig.~\ref{Fig:Minkowski}, and the rest-frame constitutive relations are then assumed to be known inputs of the problem. Although it was only first applied to plasmas in the mid 1960s~\cite{Tai1965}, this method known as Minkowski's theory had long been used and become the standard theory to model waves in moving isotropic media~\cite{Sommerfeld1952}. Compared to the other methods, Minkowski's theory lends itself more easily to relativistic flows~\cite{Chen1966,Lee1966} and covariant models ~\cite{Melrose1973,Hebenstreit1979}. On the other hand the equivalent constitutive relations, i.~e. as shown in Fig.~\ref{Fig:Minkowski} the relations between the fields in the laboratory frame obtained through Lorentz-transformation of the fields in the known rest-frame constitutive relations, are generally very intricate for a magnetized plasma. Specifically, a non-uniformly moving magnetized plasma appears as an inhomogeneous bi-anisotropic medium~\cite{Kong2008}. Deriving a wave equation is then only possible for very specific configurations. 

Recognizing this challenge, a proposed alternative is to carry out the Lorentz-transformation directly in the rest-frame dispersion relation $\mathcal{D}'$~\cite{Scarf1961}, which is obtained through the combination of the known rest-frame constitutive relations and Maxwell's equations, to get as shown in Fig.~\ref{Fig:Minkowski} the lab-frame dispersion relation $\mathcal{D}$. This relies on the on the Lorentz-covariance of dispersion relations~\cite{Censor1980,Censor2010}. This dispersion relation covariance scheme was then applied to cold magnetized plasmas~\cite{Unz1966a,Chawla1967} for which the rest-frame dispersion relation $\mathcal{D}'$ is the well-known Appleton-Hartree equation, but also to hot fluid~\cite{Unz1968} or kinetic Maxwellian plasmas~\cite{Ko1978}. Although convenient, it should be noted that the dispersion relation covariance scheme only provides the dispersion relation in the lab-frame $\mathcal{D}$, as opposed to the full wave equation. As such it cannot in itself inform on the polarization of the modes in the lab-frame. Also, another downside of the dispersion relation covariance scheme is that it assumes a specific form for Maxwell's equation in $\Sigma'$, and thus suffers from similar shortcomings as the rest-frame perturbative method introduced in the previous paragraph. 

Furthermore, while less obvious, Minkowski's theory (both in its original form and the dispersion relation covariance) still presents a number of challenges, especially for a non-uniform motion. For instance one has, in the case of an accelerated motion, to resort to the use of an instantaneous inertial rest-frame in which the medium is momentarily at rest~\cite{VanBladel1976}, from which Lorentz-transformation are then carried out. Yet, because of inertia, constitutive relations in this instantaneous inertial rest-frame do not exactly take the same form as that in the same medium at rest~\cite{VanBladel1976,Shiozawa1973}, and are therefore generally strictly speaking unknown. However, since it has been suggested that in dielectrics these corrections should be negligible if the rotation frequency remains much below the characteristic material frequencies~\cite{Shiozawa1973,Shiozawa1974}, and by lack of better alternatives, these inertia corrections have for the most part been neglected. Another basic problem of Minkowski's theory is its reliance on constitutive relations, which often are derived assuming a specific form (e.~g. plane wave or at least time-harmonic) for the perturbation. This is problematic in that the form of the solutions cannot be known a priori as it precisely depends on the constitutive relations, and imposing a specific form will affect the obtained wave equation. Thus, one should be particularly cautious if trying to derive a generic wave equation from constitutive relations. 


\section{Drag effects via Minkowski's theory}
\label{Sec:Sec_III}

Notwithstanding the issues identified above, which we will return to in the next section, let us outline what can be learned about waves in moving plasmas from Minkowski's theory. As already indicated we will write $\Sigma$ and $\Sigma'$ the laboratory-frame and the rest-frame, respectively. A prime to indicates variables written in $\Sigma'$, while unprimed variables refer to $\Sigma$ quantities. An overhead bar indicates variables characterizing a medium at rest in an inertial frame. In particular, we write
\begin{equation}
\hat{\bar{\bm{\chi}}}(\omega) = \begin{pmatrix}
\bar{\chi}_{\perp}(\omega) & -i\bar{\chi}_{\times}(\omega) & 0\\\
i\bar{\chi}_{\times}(\omega) & \bar{\chi}_{\perp}(\omega) & 0\\
0 & 0 & \bar{\chi}_{\parallel}(\omega)
\label{Eq:dielectric_tensor}
\end{pmatrix}.
\end{equation}
the usual susceptibility tensor of a static cold magnetized plasma with a background magnetic field $\mathbf{B}_0 = B_0\mathbf{\hat{e}}_{z}$, where
\begin{subequations}
\label{Eq:Tensor_component}
\begin{align}
\bar{\chi}_{\perp}(\omega) & = \sum\limits_{\alpha}\frac{{\omega_{p\alpha}}^2}{{\Omega_{c\alpha}}^2-\omega^2}\label{Eq:Tensor_component_perp}\\ 
\bar{\chi}_{\times}(\omega) & = \sum\limits_{\alpha}\epsilon_{\alpha}\frac{\Omega_{c\alpha}}{\omega}\frac{{\omega_{p\alpha}}^2}{\omega^2-{\Omega_{c\alpha}}^2}\label{Eq:Tensor_component_cross}\\
\bar{\chi}_{\parallel}(\omega) & = -\sum\limits_{\alpha}\frac{{\omega_{p\alpha}}^2}{\omega^2}\label{Eq:Tensor_component_para},
\end{align}
\end{subequations}
with $\Omega_{c\alpha} = |q_\alpha|B_0/m_\alpha$ and $\omega_{p\alpha} = [n_{\alpha} e^2/(m_{\alpha}\epsilon_0)]^{1/2}$ the unsigned cyclotron frequency and plasma frequency of species $\alpha$, respectively, and $\epsilon_{\alpha} = q_\alpha/|q_\alpha|$. Because we neglect for now inertial corrections, we further have $\hat{\bar{\bm{\chi}}}(\omega)=\hat{\bm{\chi}}'(\omega)$. The medium's velocity in the laboratory frame is $\mathbf{v}(\mathbf{x})=\bm{\beta}(\mathbf{x})c$ with $c$ the speed of light. 

\subsection{Ray deflection or transverse Fresnel drag}

\begin{figure}
\begin{center}
\includegraphics[]{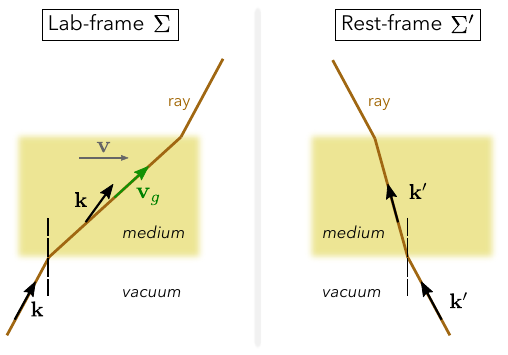}
\caption{Fresnel drag, i.e. deviation of the ray induced by motion in an isotropic medium, as seen from the lab-frame $\Sigma$ (left) and from the rest-frame $\Sigma'$ (right). Although the phase and group velocity are parallel in the rest-frame, this is not the case in the lab-frame, even if the medium is isotropic.}
\label{Fig:Fresnel}
\end{center}
\end{figure}

The change in ray direction experienced by a wave propagating through a moving media is arguably the best known manifestation of drag. As we shall show now, this result emerges naturally from the dispersion relation covariance~\cite{Censor1980,Censor2010,Carusotto2003}. To this end, consider first an isotropic dispersive dielectric medium with relative permittivity at rest $\bar{\epsilon}(\omega)$. Because we neglect inertia corrections, this is also the permittivity of this same medium in its rest-frame, i.e. $\epsilon'(\omega')=\bar{\epsilon}(\omega')$. The dispersion relation in the rest-frame $\Sigma'$ then takes its usual form
\begin{equation}
\mathcal{D}'(\omega',\mathbf{k}')=\epsilon'(\omega')\frac{\omega'^2}{c^2} - |\mathbf{k}'|^2.
\end{equation}
Now applying the dispersion relation covariance, using for simplicity the first order Lorentz-transformation of the 4-momentum vector $(\omega',\mathbf{k}')$, i.e. 
\begin{subequations}
\begin{align}
\omega' &= \omega-k_\parallel v\\
k_\parallel' &= k_\parallel-\omega v/c^2\\
\mathbf{k}_\perp' &= \mathbf{k}_\perp,
\end{align}
\end{subequations}
and finally that $\omega^2/c^2-|\mathbf{k}|^2$ is Lorentz invariant, one gets the dispersion relation in the laboratory-frame 
\begin{equation}
\label{Eq:dispersion_iso_lab_frame}
\mathcal{D}(\omega,\mathbf{k})=\frac{\omega^2}{c^2} - |\mathbf{k}|^2 +\left[\epsilon'(\omega-k_\parallel v)-1\right]\frac{(\omega-k_\parallel v)^2}{c^2}.
\end{equation}
This dispersion relation $\mathcal{D}(\omega,\mathbf{k})$ in $\Sigma$, together with the definition of the group velocity 
\begin{equation}
\mathbf{v}_g\doteq-\left(\frac{\partial\mathcal{D}(\omega,\mathbf{k})}{\partial \mathbf{\omega}}\right)^{-1}\frac{\partial \mathcal{D}(\omega,\mathbf{k})}{\partial \mathbf{k}},
\label{Eq:defintion_v_g}
\end{equation}
then show that $\mathbf{k}$ and $\mathbf{v}_g$ are not parallel in general. This result, which is illustrated in Fig.~\ref{Fig:Fresnel}, is in sharp contrast with the usual result in this isotropic medium at rest. To see this, consider for instance here the specific case of a wave whose wavevector is normal to the velocity ($\mathbf{v}\cdot\mathbf{k}=0$). From Eq.~\eqref{Eq:dispersion_iso_lab_frame} we find that in the medium the component of the group velocity along the velocity $\mathbf{v}$ is then finite even if $k_\parallel=0$, with
\begin{align}
    -\left.\frac{\partial \mathcal{D}(\omega,\mathbf{k})}{\partial k_\parallel}\right|_{k_\parallel=0} & = 2\beta\frac{\omega}{c}\left[\epsilon'(\omega)+\frac{\omega}{2}\frac{\partial \epsilon}{\partial\omega}-1\right]\nonumber\\
    & = 2 \beta \frac{n'(\omega)\omega}{c}\left[n_g'(\omega)-\frac{1}{n'(\omega)}\right].\label{Eq:drag_iso}
\end{align}
The group velocity hence has a component along the velocity $\mathbf{v}$ even if $k_\parallel=0$, that is even if the wavector is normal to the velocity. This is Fresnel drag, and we in fact recognize in Eq.~\eqref{Eq:drag_iso} the usual drag coefficient for dispersive media $(n_g'-1/n')$, as already found in Eq.~\eqref{Eq:scaling}.

Fresnel transverse drag is often considered in the context of a wave incident on a moving medium as shown in Fig.~\ref{Fig:Fresnel}, and more particularly through the transverse displacement experienced by the wave as it passes through this moving medium~\cite{Jones1972,Player1975,Carusotto2003,Banerjee2022}. For a moving isotropic medium, this transverse displacement can be obtained by considering as originally done by Player\cite{Player1975} refraction in the medium rest-frame $\Sigma'$ (right panel in Fig.~\ref{Fig:Fresnel}), taking advantage that group and phase velocity are parallel in this frame. Another option is to compute directly from generalized Snell's laws valid in the laboratory-frame the transmitted wavevector~\cite{Langlois2025}, and then compute from the laboratory frame dispersion relation and Eq.~\eqref{Eq:defintion_v_g} the group velocity for this specific wavector. 

Naturally, transverse Fresnel drag also applies to waves in moving anisotropic media such as magnetized plasmas. Yet, the extension of the derivation of Fresnel drag to this case calls for a couple of comments. First, because an anisotropic medium can support multiple modes that satisfy different dispersion relations, the drag has to be considered on a mode by mode basis. Second, if using rest-frame approach, one cannot use the result that group and phase velocity are parallel in this frame since the medium is now anisotropic in its rest-frame.

Avoiding for a moment the second difficulty, one may take advantage that certain modes in anisotropic media behave as if isotropic, i~e. that the wave index of the said mode is independent of the propagation direction $\mathbf{k}$. In a cold magnetized plasma with susceptibility tensor Eq.~\eqref{Eq:dielectric_tensor} this is notably the case of the O and X modes propagating perpendicular to the static magnetic field $\mathbf{B}_0$~\cite{Ginzburg1964}. Considering the effect of motion on these modes, and in particular of a velocity $\mathbf{v}$ that is also in the plane normal to $\mathbf{B}_0$ so that $\mathbf{k'}$ remains normal to $\mathbf{B}_0$, it was then shown that in this particular case drag verifies the usual isotropic scaling $(n_g'-1/n')$, although this scaling is now mode dependent~\cite{Langlois2025}.

An interesting consequence of this result is that the O mode is not affected by motion. Indeed, because the wave index of the O mode writes 
\begin{equation}
n' = \sqrt{1-\sum_\alpha\left(\frac{\omega_{p\alpha}}{\omega}\right)^2},
\end{equation}
the condition $n_g' n'=1$ for which transverse drag remarkably vanishes~\cite{Arnaud1976} is verified. This result is consistent with the finding that a moving cold unmagnetized plasma does not bend light~\cite{Ko1978}. Another way to get to this conclusion is to use the dispersion relation covariance. Specifically, one finds that the two modes supported by the moving plasma verify dispersion relations that are simply the Lorentz transform of dispersion functions $\mathcal{D}'_O(\omega',\mathbf{k'})$ and $\mathcal{D}'_X(\omega',\mathbf{k'})$ of the usual modes in the rest-frame $\Sigma'$~\cite{Ko1978,Mukherjee1975,MeyerVernet1980}. Then, since $\mathcal{D}'_O(\omega',\mathbf{k'})$ is singularly Lorentz invariant~\cite{Ko1978,Mukherjee1975,MeyerVernet1980}, the dispersion relation of the O mode in $\Sigma$ is independent of motion, implying from Eq.~\eqref{Eq:defintion_v_g} that there cannot be drag. This is on the other hand not the case of the X mode, for which as showed by Meyer-Vernet~\cite{MeyerVernet1980} the rays are deflected by motion. Although drag effects are generally limited for waves with large group velocity, as is typically the case of high frequency plasma waves, it has been more recently suggested that they could become more important for slower waves such as Alfv{\'e}n waves~\cite{Langlois2025}. This is reminiscent of the augmented effects observed in isotropic dielectrics under slow light conditions~\cite{Franke-Arnold2011}. Because slow waves can exist naturally in plasmas, this suggests non negligible drag effects could occur for such plasma waves.

Going back to the more general case where the group and phase velocity are not parallel in the rest-frame $\Sigma'$ a recently proposed workaround to compute drag is to use the formal definition of the group velocity Eq.~\eqref{Eq:defintion_v_g}. Indeed, using the inverse Lorentz-transformation for the 4-momentum vector, the beam drag can be formally written as a function of the rest-frame wavevector $\mathbf{k}$ and wave frequency $\omega$, which can be fully determined given the rest-frame dispersion relation $\mathcal{D}'$, the velocity $\mathbf{v}$ and the incident 4-moment vector $(\omega,\mathbf{k})$~\cite{Langlois2025}. Equipped with these tools, it becomes possible to quantify drag effects in a generic manner, and it has for instance been shown that drag effects and rest-frame anisotropy can compete with one another~\cite{Langlois2025}. These results can be used further to model drag in nonuniformly moving anisotropic media, and particular plasmas~\cite{Langlois2025a}.

\subsection{Polarization drag}
\label{Sec:polarization_drag}

Another well established manifestation of motion on wave propagation is the polarization drag experienced by a wave propagating through a rotating isotropic dielectrics, as first conjectured by Thomson~\cite{Thomson1885} and Fermi~\cite{Fermi1923}, and later demonstrated experimentally by Jones using a rotating glass rod~\cite{Jones1976}. 

\begin{figure}
\begin{center}
\includegraphics[width=9.cm]{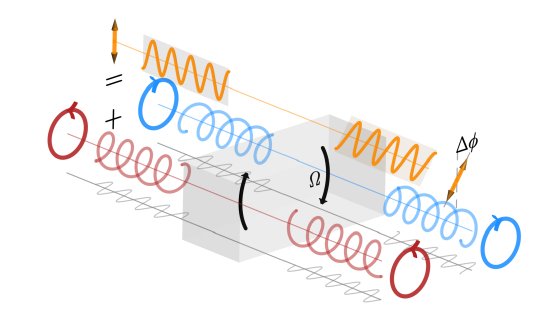}
\caption{Polarization drag $\Delta \phi$ experienced by a linearly polarized wave (orange) upon propagating along the rotation axis of a rotating dielectrics (gray box). The linearly polarized wave can be decomposed on two circularly polarized eigen-modes (red and blue) with equal amplitude. Polarization rotation results from the phase-shift between these eigenmodes introduced by rotation  (see.~\eqref{Eq:index_square_rotating_isotropic}). }
\label{Fig:SAM}
\end{center}
\end{figure}

Let us recall here how this phenomenon can be explained using Minkowski's theory introduced above, as first done by Player~\cite{Player1976}. Specifically, starting from the simple rest-frame constitutive relations for an isotropic medium
\begin{subequations}
\begin{gather}
\mathbf{D}'=\epsilon_0\left[1+\chi'(\omega')\right]\mathbf{E}\\
\mathbf{B}'=\mu_0\mathbf{H}',
\label{Eq:const_iso_sigma'}
\end{gather}
\end{subequations}
using the Lorentz-transformation for the electromagnetic fields for a medium moving with velocity $\mathbf{v}=\bm{\Omega}\times\mathbf{r}$ with $\Omega$ constant (\emph{i.~e.} solid body rotation), and Maxwell's equations in the laboratory frame, one obtains for the particular case of a plane wave with $\mathbf{k}\parallel\bm\Omega$ the wave equation for the electric displacement $\mathbf{D}$ in $\Sigma$~\cite{Player1976}
\begin{equation}
\label{Eq:wave_eq_iso}
-c^2\bm{\nabla}^2\mathbf{D}=-\left[1+\chi'(\omega')\right]\frac{\partial^2}{\partial t^2}\mathbf{D}+2\chi'(\omega')\bm{\Omega}\times\frac{\partial}{\partial t}\mathbf{D},
\end{equation}
valid to first order in $\beta=|\mathbf{v}|/c$. It can be shown that the magnetic field $\textbf{B}$ verifies the same equation. Solutions of this wave equation are circularly polarized eigenmodes, with indexes 
\begin{subequations}
\label{Eq:index_square_rotating_isotropic}
\begin{gather}
{n_{rcp}}^{2} =  1+{\chi}'(\omega-\Omega)-2\frac{\Omega}{\omega}{\chi}'(\omega-\Omega)\\
{n_{lcp}}^{2} =  1+{\chi}'(\omega+\Omega)+2\frac{\Omega}{\omega}{\chi}'(\omega+\Omega).
\end{gather} 
\end{subequations}
Because ${n_{rcp}}(\omega)\neq{n_{lcp}}(\omega)$, the polarization of a linearly polarized wave constructed as the sum of left- and right-circularly polarized modes of equal amplitude will rotate as it propagates along the rotation axis of a rotating dielectrics. This phenomenon, which is illustrated in Fig.~\ref{Fig:SAM}, is a manifestation of circular birefringence, induced here by rotation. Because circularly polarized modes correspond to spin angular momentum $\pm\hbar$, this is an effect of rotation on the spin angular momentum component of the wave. Quantitatively, the rotation angle over a distance $l$ is 
\begin{equation}
\Delta \phi = [{n_{lcp}}-{n_{rcp}}]\frac{\omega l}{2c},
\label{Eq:polarization_rotation_def}
\end{equation} 
which, using Eq.~\eqref{Eq:index_square_rotating_isotropic}, writes \begin{equation}
\label{Eq:Player_polarization_drag}
\Delta \phi=\frac{\Omega l}{c}[n_g'(\omega)-n^{-1}(\omega)]
\end{equation} 
to lowest order in $\Omega$. Here we once more recognize the Fresnel drag coefficient  $(n_g'-1/n')$.

This rotation induced birefringence bears close similarities with the classical Faraday rotation induced by a magnetic field~\cite{Faraday1846}. Specifically, for a wave propagating along the background magnetic field $\mathbf{B}_0$ in a static cold magnetized plasma with susceptibility tensor Eq.~\eqref{Eq:dielectric_tensor}, the wave indexes write~\cite{Ginzburg1964} 
\begin{subequations}
\label{Eq:index_square_rotating_magplasma}
\begin{gather}
\bar{n}_{rcp}^{2} =  1+\bar{\chi}(\omega)+\bar{\chi}_{\times}(\omega)\\
\bar{n}_{lcp}^{2} =  1+\bar{\chi}(\omega)-\bar{\chi}_{\times}(\omega).
\end{gather} 
\end{subequations}
Eq.~\eqref{Eq:polarization_rotation_def} shows that the off-diagonal terms $\bar{\chi}_\times$ induced by the magnetic field similarly lead to a phase shift between left-and right-circularly polarized eigenmodes. This manifests as a polarization rotation, with in the limit of high frequency waves $\omega\gg\omega_{pe}$ and $\omega\gg\Omega_{ce}$ a rotation angle 
\begin{equation}
\label{Eq:Player_polarization_drag}
\Delta \phi\sim\frac{\omega l}{2c}\bar{\chi}_{\times}\sim\frac{\Omega_{ce}\omega_{pe}^2l}{2c\omega^2}.
\end{equation}
Comparing Eq.~\eqref{Eq:index_square_rotating_magplasma} with Eq.~\eqref{Eq:index_square_rotating_isotropic} shows that the magnetic term $\bar{\chi}_{\times}$ plays a similar role as the mechanical term $2\Omega/\omega\bar{\chi}'$. This can be interpreted as a manifestation of Larmor's theorem and the equivalence between the Coriolis force in a rotating frame and the magnetic force. 


Building on these findings in isotropic media, it was then shown that the wave equation Eq.~\eqref{Eq:wave_eq_iso} could in fact be extended to the case of a rotating gyrotropic media~\cite{Gueroult2019a}, and thus to magnetized plasmas. Specifically, considering the particular case of a cold rotating magnetized plasma for which the magnetic field is parallel to the rotation axis, referred to as an aligned rotator, it was shown that the eigenmodes remain circularly polarized with, to lowest order in $\Omega$, with indexes
\begin{subequations}
\label{Eq:index_general}
\begin{multline}
{n_{rcp}}^2(\omega)  = 1+ {\chi}'_{\perp}(\omega-\Omega) + {\chi}'_{\times}(\omega-\Omega) \\-\frac{\Omega}{\omega}\left[{\chi}'_{\times}(\omega-\Omega) + {\chi}'_{\parallel}(\omega-\Omega)+{\chi}'_{\perp}(\omega-\Omega)\right],\label{Eq:index_general_r}
\end{multline}
and
\begin{multline}
{n_{lcp}}^2(\omega) = 1+ {\chi}'_{\perp}(\omega+\Omega) - {\chi}'_{\times}(\omega+\Omega) \\-\frac{\Omega}{\omega}\left[{\chi}'_{\times}(\omega+\Omega) - {\chi}'_{\parallel}(\omega+\Omega)-{\chi}'_{\perp}(\omega+\Omega)\right].
\label{Eq:index_general_l}
\end{multline}
\end{subequations} 
Eq.~\eqref{Eq:index_general} reduces as expected to Eq.~\eqref{Eq:index_square_rotating_isotropic} in the limit of an isotropic medium.

Examining the solutions in Eq.~\eqref{Eq:index_general} shows that polarization rotation then occurs through a combination of the two effects introduced above. One is the Faraday rotation~\cite{Faraday1846} associated with the finite rest-frame off-diagonal susceptibility component ${\chi}'_{\times}$. It however here is computed for the Doppler shifted wave frequency $\omega'$. The other, which corresponds to the bracketed terms in Eqs.~\eqref{Eq:index_general_r} and \eqref{Eq:index_general_l}, is, as shown by the $\Omega$ dependence, a mechanical contribution due to the plasma rotation, \emph{i.~e.} polarization drag. Depending on whether $\bm{\Omega}$ and $\mathbf{B}$ are parallel or anti-parallel, polarization drag now competes with or supplements Faraday rotation. 

Trying to make sense of these results, Eqs.~\eqref{Eq:index_general_r} and \eqref{Eq:index_general_l} show that one of the two modes cannot propagate when the bracketed term becomes larger than the sum of the other terms on the right hand side. For a magnetized plasma this necessarily happens for sufficiently low frequency since, from Eq.~\eqref{Eq:Tensor_component_para}, ${\chi}'_{\parallel}(\omega)$ diverges when $\omega\rightarrow 0$. There is accordingly a low frequency cutoff below the ion-cyclotron frequency below which one of the two modes does not propagate. Interestingly, while Faraday rotation does generally dominate over polarization drag, polarization drag can in fact be order of magnitude larger than Faraday rotation in a specific frequency band above this cutoff~\cite{Gueroult2020}. 

Because this effect is non-reciprocal, that is to say that a wave propagating back and forth along the plasma column will experience double the polarization rotation, and because of the weak losses found in plasmas, this ability to control polarization through rotation could be used to design agile polarization manipulation components such as isolators and circulators~\cite{Gueroult2020}.  Another promising setting where polarization drag could play a role is electron-positron plasmas, since in that case Faraday rotation cancels out. Building on this observation, it was suggested that polarization drag could play an important role in pulsars' magnetosphere~\cite{Gueroult2019a}, and from there on the estimation of interstellar magnetic fields using pulsar polarimetry.

\subsection{Image rotation}

A third manifestation of rotation, closely related to the first two, is the phenomenon known as image rotation~\cite{Padgett2006}, that is a rotation of the transverse structure of a wave propagating through a rotating medium. 

\begin{figure}
\begin{center}
\includegraphics[width=9.cm]{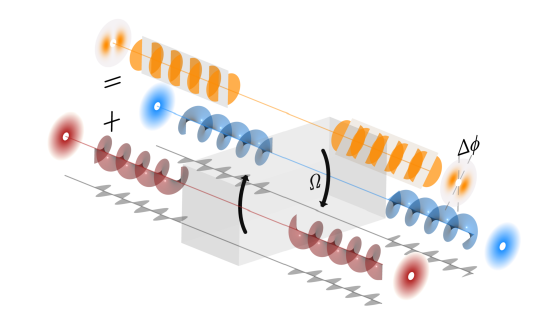}
\caption{Image rotation $\Delta \phi$ experienced by a wave with transverse structure (orange) upon propagating along the rotation axis of a rotating dielectrics (gray box). The two lobes structure can be decomposed on two orbital angular momentum (OAM) eigenmodes of the form $J_l(\alpha r)\exp[i(l\theta+k_z z)]$ (red and blue) with equal amplitude. Image rotation results from the phase-shift between these OAM eigenmodes introduced by rotation. }
\label{Fig:OAM}
\end{center}
\end{figure}

Formally, this effect can be similarly derived from Minkowski's theory by extending Player's original derivation~\cite{Player1976} to more complex wave forms. This was notably first done by G{\"o}tte \emph{et al.} who showed considering the same isotropic dielectric Eq.~\eqref{Eq:const_iso_sigma'} in solid body rotation that retiring the assumption of a plane wave 
then led to the wave equation~\cite{Goette2007} 
\begin{multline}
\label{Eq:wave_eq_iso_image_rotation}
-c^2\bm{\nabla}^2\mathbf{D}=-\left[1+\chi'(\omega')\right]\frac{\partial^2}{\partial t^2}\mathbf{D}\\+2\chi'(\omega')\left[\bm{\Omega}\times\frac{\partial}{\partial t}\mathbf{D}-\Omega\frac{\partial^2}{\partial\theta\partial t}\mathbf{D}\right].
\end{multline}
Compared to Eq.~\eqref{Eq:wave_eq_iso}, Eq.~\eqref{Eq:wave_eq_iso_image_rotation} contains an additional term due to the now finite $(\mathbf{v}\cdot\bm{\nabla})\partial\mathbf{D}/\partial t$. This term is responsible for image rotation, whereas the first term in brackets on the second line of Eq.~\eqref{Eq:wave_eq_iso_image_rotation} corresponds to polarization drag. Considering specifically waves with a phase of the form $\exp[i(l\theta+k_z z-\omega t)]$, that is wave with an orbital angular momentum (OAM) $\hbar l$, G{\"o}tte \emph{et al.} showed that rotation introduces a phase shift between modes with opposite OAM $\pm l$. Just like a phase shift between spin angular momentum (SAM), i.~e. circularly polarized modes, led to polarization rotation, a phase shift between OAM modes with opposite OAM $\pm l$ now leads to a rotation of the wave transverse structure. This phenomenon is illustrated in Fig.~\ref{Fig:OAM}. For an isotropic dielectric the angle of polarization rotation is in fact equal to the angle of image rotation, confirming how rotation acts similarly on the spin and orbital components of the wave.~\cite{Padgett2006,Goette2007,WisniewskiBarker2014}. 

Although extending G{\"o}tte \emph{et al.}'s derivation of image rotation using Minkowski's theory to anisotropic media such as magnetized plasmas poses significant mathematical challenges and, to our knowledge, has not been achieved to date, the possibility for image rotation in plasmas has been identified through different means. Specifically, that plasmas can support waves carrying OAM is now well established, both in unmagnetized~\cite{Mendonca2009,Mendonca2012,Mendonca2012a,Chen2017} and magnetized~\cite{Shukla2012,Stenzel2015,Urrutia2016,Stenzel2016} plasmas. In a static plasma, however, the dispersion relation satisfied by modes with opposite OAM $\pm l$ is the same, i.~e. $k_z (\omega,l)=k_z (\omega,-l)$, so that there is no phase shift between the modes, and thus no image rotation when considering the combination of these $\pm l$ modes. Following a similar approach but considering now perturbations around a rotating equilibrium, it was recently shown that this was no longer the case for modes in a rotating plasma. In particular, it was found considering more specifically Trivelpiece-Gould, Whistler-Helicon and Alfv{\'en} modes in a rotating plasma that the axial wavevector of modes with opposite OAM differ~\cite{Rax2021,Rax2023b}, i.~e. $k_z (\omega,l, \Omega)\neq k_z (\omega,-l,\Omega)$. As a consequence the structure of the sum of these modes $\pm l$ will rotate along the $z$-axis, suggesting that image rotation also manifests for plasma waves. 

In fact, these predictions were recently demonstrated experimentally in the Large Plasma Device (LAPD)~\cite{Gekelman2016} at the University of California Los Angeles. This was made possible through the combination of a controlled rotation of the plasma column enabled by segmented rings end-electrodes biased at different potentials~\cite{Gueroult2024}, and the low group velocity of shear Alfv{\'e}n waves routinely studied in the machine~\cite{Gekelman2011}. The observed transverse structure rotation~\cite{Gueroult2025} was then found to match with good accuracy the transverse Fresnel drag expected for the specific operating conditions, and the image rotation predicted by theory~\cite{Rax2023b}.

To conclude this section, we must emphasize that the discussion of drag effects given above has been conducted assuming a cold magnetized plasma, with dielectric tensor given by Eqs.~\eqref{Eq:dielectric_tensor} and \eqref{Eq:Tensor_component}. Although convenient to expose basic effects, this assumption is rarely valid in real applications, with finite temperature often affecting the plasma's behaviour. Under this light, one may ask how finite-temperature corrections will affect the drag effects exposed here, or even introduce new ones. For example, due to finite-temperature corrections, a wave propagating in a warm unmagnetized plasma no longer verifies $n_g'n'\neq1$, and therefore drag effects no longer vanish. Although this problem remains to be addressed in full, preliminary results suggest thermal corrections will become noticeable when the thermal speed is comparable to the wave phase velocity.

\section{Outstanding questions and prospects}
\label{Sec:Sec_IV}

As already hinted at in Section~\ref{Sec:Sec_II}, the use of Minkowski's theory to model propagation in media in non-uniform motion poses a number of basic questions. Identifying these challenges and possible workarounds is important to accurately model motion effects in more complex configurations. Analyzing and confronting the results and approaches of previous studies, we now underline and discuss a number of key issues.

\subsection{Inertia corrections}

A first difficulty with using Minkowski's theory to model propagation in moving media arises when considering the effect of a non-uniform motion. The root of this difficulty lies in the reliance of Minkowski's theory on an \emph{instantaneous rest-frame} or co-moving frame~\cite{VanBladel1976} - which we recall is defined as the inertial frame in which the medium is instantaneously and locally at rest - from which Lorentz-
transformation can then be carried out. Specifically, Minkowski's theory demands the constitutive equations in this frame. Formally, the difficulty lies in the fact that for a non-uniform motion an observer at rest in this inertial frame experiences a finite acceleration and sees a curved space-time. 

However, writing $\mathbf{a}$ the acceleration, this frame can be considered locally flat over distances $d$ such that $|\mathbf{a}|d\ll c^2$~\cite{Misner2017}. Considering more specifically  a rotation at angular frequency $\bm{\Omega}$, this corresponds simply to $\Omega d\ll c$, that is non-relativistic flows. In this limit, Maxwell's equations then retain their conventional form but inertia corrections appear in the form of modified constitutive equations~\cite{Steinberg2023}, which differ from those classically known to hold in this same medium at rest~\cite{Anderson1969,Shiozawa1973,VanBladel1976}. To give a practical example, while $\mathbf{D}'=\epsilon_0\left[\hat{\mathds{1}}+\hat{\bar{\bm{\chi}}}(\omega')\right]\mathbf{E}'$ holds in a cold magnetized plasma at rest or in uniform motion, this is no longer the case if the plasma is in non-uniform motion, and one needs to express $\hat{{\bm{\chi}'}}(\omega')$. 

To go beyond simply acknowledging the role of inertia in $\Sigma'$, one needs to consider a particular velocity field. In light of previous discussions and experimental interest, a candidate of choice is rotation. In this particular case, and considering more specifically the effect on the dielectric response of the medium, it has been shown from phenomenological models that a medium that is characterized by an isotropic response when at rest, that is $\hat{\bar{\chi}} = \bar{\chi}\hat{\mathds{1}}$, appears gyrotropic in its rest-frame when in rotation~\cite{Shiozawa1973,Shiozawa1974,Langlois2023}. These corrections are due to the the Coriolis force that manifests in the rotating rest-frame.

Considering now the possible effect of these corrections on wave observables, it was suggested soon after Jones' demonstration of polarization drag~\cite{Jones1976} that Player's derivation~\cite{Player1976}, which assumed rigid rotation, should be corrected to account for inertia~\cite{Baranova1979}. Specifically, the idea then proposed by Baranova and Zel'dovich~\cite{Baranova1979} to model inertia effects on polarization drag relied on Larmor's theorem, and from there on a Faraday-Coriolis equivalence. This approach however demands a model for Faraday rotation, for instance through the Verdet constant, which may not always be known. It can also by construction only model inertia effect on one species, prohibiting its use in plasmas. Addressing this point, it was more recently shown that inertia corrections to polarization drag in a magnetized plasma are negligible~\cite{Langlois2023}, at least in the frequency range where polarization drag dominates over Faraday rotation~\cite{Gueroult2019a,Gueroult2020}. Exploring further these effects, it was also shown that inertia can, on the other hand, play a key role on polarization drag in a rotating cold unmagnetized plasma~\cite{Langlois2024}. In fact, because of the singular property that, as already discussed in the context of Fresnel drag, $\bar{n}_g \bar{n}=1$ in an unmagnetized plasma, Player's result Eq.~\eqref{Eq:Player_polarization_drag} predicts zero polarization rotation to first order in $\beta$. Yet, by considering inertia corrections to the rest-frame dielectric tensor of this rotating unmagnetized plasma, it was shown that there is a non-zero polarization drag
\begin{equation}
\Delta \phi^{up} = \frac{\Omega l}{c}\left[\bar{n}(\omega)-\frac{1}{\bar{n}(\omega)}\right].
\label{Eq:pola_rotation_up}
\end{equation}
This contribution then comes entirely from inertia. This finding is interesting not only in that it gives a means to probe rotating plasmas, but also because a measure of polarization drag in such conditions could conversely provide a demonstration of the role of inertia corrections, which as mentioned earlier have so far received little attention.

Looking ahead, a question that remains is how to account for inertia in more generic configurations (that is velocity fields), for which simple rest-frame model such as those obtained for rotation do not exist. Addressing this point is key to properly accounting for inertia when modeling waves in moving media through Minkowski's theory~\cite{Anderson1969}, since as mentioned above this approach demands the constitutive relations in the instantaneous rest-frame. 

\subsection{Limits of plane wave models}

Another basic question that arises in the modeling of wave propagation in moving media, and in particular in plasmas, is that of the nature of solutions. More specifically, plane wave solutions in general only hold for homogeneous media, which seems contradictory with media exhibiting non-uniform motion. 

To see this, recall that from the usual geometrical optics ray equations~\cite{Born2019,Tracy2014} the evolution of the wavector $\mathbf{k}$ along the ray is given by 
\begin{equation}
\frac{d\mathbf{k}}{ds} = \frac{\partial \mathcal{D}(\omega,\mathbf{k},\mathbf{x})}{\partial \mathbf{x}}
\end{equation}
Now, using the dispersion relation covariance $\mathcal{D}(\omega,\mathbf{k},\mathbf{x})=\mathcal{D'}(\omega(\omega',\mathbf{k}'),\mathbf{k}(\omega',\mathbf{k}'),\mathbf{x})$, the Lorentz-transformation for the 4-momentum vector and the chain rule, this  formally gives to first order in $\beta$
\begin{align}
\frac{d\mathbf{k}}{ds} = &\frac{\partial \mathcal{D'}(\omega',\mathbf{k'},\mathbf{x})}{\partial \mathbf{x}}\nonumber\\
&-c\frac{\partial \left[\bm{\beta}(\mathbf{x})\cdot\mathbf{k}\right]}{\partial \mathbf{x}}\frac{\partial \mathcal{D'}(\omega',\mathbf{k'},\mathbf{x})}{\partial \omega'}\nonumber\\
&-\frac{\omega}{c}\frac{\partial \bm{\beta}(\mathbf{x})}{\partial \mathbf{x}}\frac{\partial \mathcal{D'}(\omega',\mathbf{k'},\mathbf{x})}{\partial \mathbf{k'}}.
\label{Eq:wavector_dev}
\end{align}
This result explicitly demonstrates that the wavector can classically be redirected from inhomogeneities in the rest-frame properties (first line in Eq.~\eqref{Eq:wavector_dev}), but also now via the non-uniformity of the medium motion $\bm{\beta}(\mathbf{x})$ (second and third lines in Eq.~\eqref{Eq:wavector_dev})~\cite{Braud2025}. Put differently, the non-uniformity  of the flow introduces a spatial dependence of the dispersion relation through the Lorentz-transformation of the 4-momentum vector, which then leads to a non-uniform $\mathbf{k}$. This general result thus shows that plane waves cannot properly describe wave propagation in non-uniformly moving media. 

\begin{figure}
\begin{center}
\includegraphics[width=8.6cm]{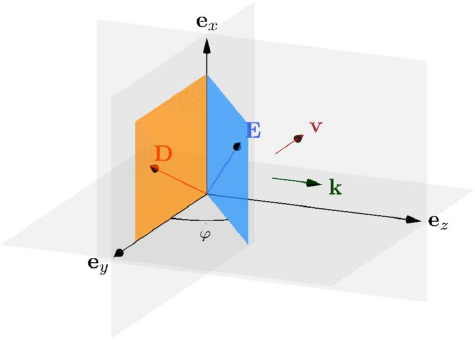}
\caption{Redirection of the oscillation plane for the electric field $\mathbf{E}$ in a moving medium. Although $\mathbf{D}$ and $\mathbf{B}$ are transverse at all points, the local velocity $\mathbf{v}(\mathbf{x})$ is responsible for a redirection of the plane of polarization for $\mathbf{E}$ (here a rotation around $\mathbf{e}_x$ since velocity $\mathbf{v}=v\mathbf{e}_y$ and $\mathbf{k}=k\mathbf{e}_z$), which is then no longer transverse. For a non-uniformly moving medium $\mathbf{v}(\mathbf{x})$, this redirection leads to an orientation which varies from point to point, so that $\mathbf{E}$ cannot be represented by a plane wave. Note that in general $\mathbf{k}$ varies spatially, so that $\mathbf{D}$ and $\mathbf{B}$ cannot be represented by a plane wave either. }
\label{Fig:PolarizationPlane}
\end{center}
\end{figure}

At first glance, this statement appears contradictory to the models presented in Section~\ref{Sec:Sec_III}, and in particular to Player's derivation of polarization drag in a rotating dielectric Eq.~\eqref{Eq:Player_polarization_drag}. This apparent contradiction is however lifted when realizing that in the singular case of interest, that is a wave normally incident on a rotating cylinder, the wavector $\mathbf{k}$ remains constant and parallel to the rotation axis. In this particular case, the oscillation plane for the $\mathbf{D}$ and $\mathbf{B}$ fields, which by definition must be normal to $\mathbf{k}$, remain normal to the rotation axis at all points, and the wave envelope is accordingly constant. This explains how a plane wave treatment~\cite{Player1976,Goette2007,Gueroult2019a,Gueroult2020,Langlois2024}, can give a polarization rotation result consistent with experiments~\cite{Jones1976} in this particular case. It is important to stress here though that, even if $\mathbf{k}$ is constant, this fortuitous result for polarization rotation is only valid when working with $\mathbf{D}$ or $\mathbf{B}$. It is, in particular, not true if working with $\mathbf{E}$, as it is usually the case in plasmas. As we will demonstrate below, this is because a moving medium appears anisotropic as seen in the laboratory-frame~\cite{Kong2008}, and the electric field $\mathbf{E}$ is thus inclined with respect to $\mathbf{k}$, with an orientation that depends on the local velocity $\mathbf{v}(\mathbf{x})$. This behavior is illustrated in Fig.~\ref{Fig:PolarizationPlane}.

Before exposing this problem, let us illustrate first how this subtle difference can lead to confusion and errors. As shown in Appendix~\ref{App:WaveE}, following Minkowski's theory in a fashion similar to that used by Player~\cite{Player1976} but working with $\mathbf{E}$, one may obtain the wave equation
\begin{equation}
c^2\bm{\nabla}\times\bm{\nabla}\times\mathbf{E} =-\left[1+\chi'(\omega')\right]\frac{\partial^2}{\partial t^2}\mathbf{E}+\chi'(\omega')\bm{\Omega}\times\frac{\partial}{\partial t}\mathbf{E}.
\label{Eq:WaveEqE}
\end{equation}
From there, assuming a harmonic time dependence $e^{-i\omega t}$, we can define the equivalent dielectric tensor 
\begin{equation}
\hat{\bar{\bm{\chi}}}_E(\omega) = \begin{pmatrix}
\chi'(\omega') & i\Omega\chi'(\omega')\Big/\omega  & 0\\
-i\Omega\chi'(\omega')\Big/\omega  & \chi'(\omega') & 0\\
0 & 0 & \bar{\chi}(\omega')
\end{pmatrix}
\label{Eq:dielectric_tensor_E}
\end{equation}
to have Eq.~\eqref{Eq:WaveEqE} take the usual form
\begin{equation}
\bm{\nabla}\times\bm{\nabla}\times\mathbf{E} =\left[\hat{\mathds{1}}+\hat{\bar{\bm{\chi}}}_E(\omega')\right]\frac{\omega^2}{c^2}\mathbf{E}.
\label{Eq:WaveEqE2}
\end{equation}
If now assuming a plane wave solution for $\mathbf{E}$, one finds through identification of the off-diagonal terms in the susceptibility tensor Eq.~\eqref{Eq:dielectric_tensor_E} with the usual off-diagonal terms of the magnetized plasma susceptibility tensor Eq.~\eqref{Eq:dielectric_tensor} that the rotation contribution is $\Omega/\omega\bar{\chi}'$, that is only half of what was found in Eq.~\eqref{Eq:index_square_rotating_isotropic} working with $\mathbf{D}$. Accordingly, one would also conclude that the polarization drag angle is only half of Eq.~\eqref{Eq:Player_polarization_drag} as obtained by Player~\cite{Player1976}. 

The cause of this error is that, as mentioned above, the orientation of the oscillation plane for $\mathbf{E}$ varies spatially with $\mathbf{v}(\mathbf{x})$. Let us now expose how this occurs. First, recall that Minkoswki's laboratory-frame constitutive relations for an isotropic dielectric with rest-frame permittivity $\chi'(\omega')$ moving with a velocity $\mathbf{v}=\bm{\beta}c$ write to first order in $\beta$~\cite{Minkowski1908}
\begin{subequations}
\label{Eq:Mink}
\begin{gather}
\mathbf{D} = \epsilon_0[1+\chi'(\omega')]\cdot\mathbf{E}+c^{-1}\chi'(\omega')\bm{\beta}(\mathbf{x})\times\mathbf{H},\\
\mathbf{B} = {\mu_0}\mathbf{H}-c^{-1}\chi'(\omega')\bm{\beta}(\mathbf{x})\times\mathbf{E}.
\end{gather}
\end{subequations}
Using Maxwell's equation, Eqs.~\eqref{Eq:Mink} can be rewritten as a pair of coupled equations for $\mathbf{E}$ and $\mathbf{H}$ only
\begin{subequations}
    \begin{gather}\label{eq:coupled_eq_curlE}
        \bm\nabla\times\mathbf{E}=-\mu_0\partial_t\mathbf{H} +\chi'(\omega')c^{-1}\bm\beta(\mathbf{x})\times\partial_t\mathbf{E}\\\label{eq:coupled_eq_curlH}
        \bm\nabla\times\mathbf{H} =\epsilon_0\left[1+\chi'(\omega')\right]\partial_t\mathbf{E} +\chi'(\omega')c^{-1}\bm\beta(\mathbf{x})\times\partial_t\mathbf{H}.
    \end{gather}
\end{subequations}
Defining 
\begin{equation}\label{Eq:def_omega}
    \mathbf{w}(\mathbf{x},\mathbf{k},\omega)=\mathbf{k}+\chi'(\omega')\frac{\omega}{c}\bm{\beta}(\mathbf{x}),
\end{equation}
the second equation Eq.~\eqref{eq:coupled_eq_curlH} then clearly shows that, 
$(\mathbf{w}, \mathbf{E}, \mathbf{H})$ forms a direct triad~\cite{Rozanov2005}, i.~e. that $\mathbf{E}$ lies in the plane normal to $\mathbf{w}$. Eq.~\eqref{Eq:def_omega} then explicitly shows that the plane of polarization for $\mathbf{E}$ is redirected by motion $\bm{\beta}(\mathbf{x})$, and is in particular not normal to $\mathbf{k}$. Specifically, the oscillation plane is rotated by an angle $\varphi$ around the direction given by $\mathbf{k}\times\bm{\beta}$. In the particular case $\mathbf{k}\perp\mathbf{v}$ one finds 
\begin{equation}
    \tan\varphi = \left(n'(\omega')-\frac{1}{n'(\omega')}\right)\beta (\mathbf{x})
\end{equation}
to lowest order in $\beta$, with $n'(\omega')=\sqrt{1+\chi'(\omega')}$. Going back now to our example of a wave propagating along the rotation axis of a rotating dielectric, for which we recall that $\mathbf{k}$ is singularly constant, Eq.~\eqref{Eq:def_omega} shows that the plane of polarization for $\mathbf{E}$ will vary spatially through $\bm{\beta}(\mathbf{x})$. This is consistent with the remark by Player in his original study~\cite{Player1976} that \textquote{$\mathbf{E}$ and $\mathbf{H}$ have longitudinal components of order $\beta$ whose amplitude depend on position}. Because of this inhomogeneity, $\mathbf{E}$ cannot be described by a plane-wave, $\bm{\nabla}\times\mathbf{E}\neq i\mathbf{k}\times\mathbf{E}$, and waves indexes can not be derived from Eq. (22) as usually done in homogeneous media.

In summary, while plane wave solutions for $\mathbf{D}$ and $\mathbf{B}$ exist for a wave propagating along the rotation axis of a rotating medium, from which polarization rotation can be straightforwardly derived, this is a very singular and uncommon situation. In general one should not expect plane waves to accurately describe waves in a non-uniformly moving medium, irrespective of working with $\mathbf{E}$ or $\mathbf{D}$. 

Related to the last point, it is important to stress here that one can analogously not expect monochromatic waves to be solutions if motion is the source of time variations of the laboratory-frame medium's properties, for instance if the medium's rest-frame properties are inhomogeneous.

\subsection{Extended geometrical optics for moving media}

Given, as highlighted in the previous paragraph, the inhomogeneous nature of the properties of moving media as seen from the laboratory-frame $\Sigma$, a natural idea is then to look at what can be learned from next-order corrections to geometrical optics, where the wavelength $\lambda$ is no longer assumed to be zero (or equivalently the medium to be homogeneous), but simply small compared to the characteristic scale of the medium inhomogeneities. The question then becomes how the effect of motion, just like inhomogeneities in a static medium, can be captured through the evolution of the wave envelope~\cite{Kravtsov1990,Tracy2014}.

To address this question, one may take advantage of the extended geometrical optics (XGO) theory developed by D. Ruiz and I. Dodin~\cite{Ruiz2015,Ruiz2017, Ruiz2017a,Dodin2019}. In contrast to traditional zero-th order geometrical optics which model waves in the form of rays that just like classical particles are fully described by their coordinates and momenta, XGO captures how the dynamics of a vector wave further depends on its polarization (i.~e. its spin) by calculating next order corrections to the geometrical optics dispersion operator. As such, it notably captures spin-orbit effects~\cite{Bliokh2015}, that is, spin-dependent corrections to the wave trajectory or, reciprocally, corrections to the wave polarization due to the geometrical phase accumulated along the ray. In particular, it models the effect of variation of the polarization due to a redirection of the ray direction by inhomogeneities.  

Because XGO theory is formulated for a general vector wave, it can notably be applied to the particular case of wave equations that describe waves in moving media. Since motion non-uniformity appears as inhomogeneities in the laboratory frame, XGO should in principle capture the effect of variation of the polarization due to a redirection of the ray by the non-uniform velocity field $\mathbf{v}(\mathbf{x})$. We should note here that examining the effect of motion through next order corrections in the geometrical optics expansion is not a new idea, as demonstrated by Rozanov and Sochilin for a non-dispersive isotropic media~\cite{Rozanov2005,Rozanov2006}. However, through its very generic formulation, XGO theory offers in principle the means to consider in a very systematic and consistent manner a much wider range of media including plasmas.

Although very recent, first efforts in this direction have already brought promising results. Specifically, considering first a non-dispersive medium, it was shown that a non-uniform motion is the source of a rotation of polarization, not only through the polarization drag effect discussed in Sec.~\ref{Sec:polarization_drag}, but also through a spin-redirection phase induced by motion~\cite{Braud2025a}. Specifically, a non-uniform motion is responsible for a bending of the ray, leading to non-planar rays, which then lead to non-trivial spin-redirection Berry (or geometrical) phase that affects polarization. In doing so this model captures consistently the effect of motion on the trajectory and on the polarization, and it in fact makes it possible to describe in a single model both the azimuthal Fresnel drag and the polarization drag experienced by a wave propagating along a rotating medium.

To illustrate how this theory captures motion effects and addresses the shortcomings of plane wave models, let us consider here the particular case of a moving unmagnetized plasma. Considering perturbations around an ideal moving equilibrium, one obtains as shown in Appendix~\ref{App:WaveELab} the wave equation for the electric field
\begin{align}
    \bm\nabla\times\bm\nabla\times\mathbf{E} = &\left[1-\frac{\omega_p^2}{\omega^2}\right]\frac{\omega^2}{c^2}\mathbf{E}\nonumber\\
    &+\frac{\omega_p^2}{\omega^2}\frac{i\omega}{c}\left(\bm\beta\times\bm\nabla\times\mathbf{E}+\bm\nabla\times\bm\beta\times\mathbf{E}\right)\nonumber\\
    &+2\frac{\omega_p^2}{\omega^2}\frac{i\omega}{c}\left(\bm\beta\cdot\bm\nabla+\bm\nabla\cdot\bm\beta\right)\mathbf{E}.
    \label{Eq:WaveEqUnmagnetizedPlasma}
\end{align}
Note that we here assumed a monochromatic solution since we considered for simplicity time-independent medium properties. However, because XGO theory treats temporal and spatial variations of the medium properties on an equal footing~\cite{Ruiz2015,Ruiz2017, Ruiz2017a,Dodin2019}, it could be similarly applied if a time dependence were to be retained, then searching for quasi-monochromatic waves. If focusing on the canonical case of propagation along the axis of a rigid rotator $\mathbf{k}\parallel\bm{\Omega}$ Eq.~\eqref{Eq:WaveEqUnmagnetizedPlasma} reduces to 
\begin{equation}
\bm{\nabla}\times\bm{\nabla}\times\mathbf{E} =\left[1-\frac{\omega_p^2}{\omega^2}+i\frac{\omega_p^2}{\omega^2}\frac{\bm{\Omega}}{\omega}\times\right]\frac{\omega^2}{c^2}\mathbf{E}
\label{Eq:WaveEqUnmagnetizedPlasmaAligned}
\end{equation}
with $\omega_p^2 = \omega_{pe}^2+\omega_{pi}^2$ the plasma frequency, which we verify is identical to Eq.~\eqref{Eq:WaveEqE} other than that the susceptibility $\chi'(\omega)=-\omega_p^2/\omega^2$ depends here on $\omega$ as opposed to $\omega'$. Assuming a plane wave solution for $\mathbf{E}$ would hence similarly lead to an erroneous result for the wave indexes, and, from there, for polarization drag. In contrast, XGO theory splits the dispersion operator $\bm{D}=\bm{D}_0+\bm{D}_1$, with $\bm{D}_0$ the zero-th order dispersion operator that is classically used for plane waves, and $\bm{D}_1$ the first order dispersion operator which captures the inhomogeneity of the medium. Because the last term in Eq.~\eqref{Eq:WaveEqUnmagnetizedPlasmaAligned} is proportional to $\bm{\Omega}$, that is a derivative of the medium properties, it should normally enter through $\bm{D}_1$ and not $\bm{D}_0$, and thus not affect the zero-th order solution. Enforcing this geometrical ordering and applying XGO theory to Eq.~\eqref{Eq:WaveEqUnmagnetizedPlasma}, we find that the polarization has indeed two origins. One is from $\bm{D}_1$, which contributes to half the polarization rotation. The other, which contributes for the other half, comes from variations of the eigenvectors with position and wavector. Importantly, since Eq.~\eqref{Eq:WaveEqUnmagnetizedPlasmaAligned} is only valid for a single $\mathbf{k}$ this term is lost when solving Eq.~\eqref{Eq:WaveEqUnmagnetizedPlasmaAligned} for a plane wave solution, which then predicts only half the polarization rotation. On the other hand, these two terms in the XGO theory together yield back Eq.~\eqref{Eq:pola_rotation_up}. Player's hybrid treatment where inhomogeneities (through $\bm{\Omega}$) appear in the wave equation for plane waves only works because as mentioned above the plane of polarization for $\mathbf{D}$ and $\mathbf{B}$ is constant in space (normal to $\mathbf{k}$ which is here singularly constant), so that the full polarization rotation then comes from $\bm{D}_1$.   

This simple example shows how XGO theory can be used to properly model propagation and in particular polarization effects in moving media. Although it has so far only been applied to moving non-dispersive dielectrics~\cite{Braud2025a} and moving unmagnetized plasmas~\cite{Braud2026}, the approach can in principle be generalized to more complex media, the main requirement being the derivation of an appropriate wave equation. For the reasons laid out above, notably the need for rest-frame constitutive relations and inertia effects, obtaining this wave equation from Minkowski's theory may not be suitable. If possible, a laboratory-frame perturbative method seems preferable in this regard.

\section{Summary}
\label{Sec:Sec_V}

The properties of a wave propagating through a medium are modified if this medium is moving with respect to the observer. Although these drag effects are relatively well understood in isotropic dielectrics, and in particular for a uniform motion, generic models for the effect of motion on plasma waves remain limited. This is limiting in that not only plasmas often feature flows, but also because the scaling of drag effects in isotropic dielectric suggests that enhanced effects could be at play in plasmas. In this paper, we review what is known of drag effects in plasmas, and discuss challenges and possible workarounds to derive models that could account for the effect of motion on plasma waves in realistic settings.

Building on well known results from the literature on drag effects in isotropic dielectrics, it is first shown using Minkowski's theory that motion can affect a number of observables of plasmas waves. Considering more specifically the case of a rotation, motion is notably shown to alter the wave trajectory or ray, its polarization and, for waveforms more complex than plane waves, its transverse structure. Beyond confirming that drag effects should be at play in plasmas, these results also indicate that, at least in some simpler configurations, the scaling of these effects remains the same Fresnel drag scaling $(n_g-1/n)$ as that known in isotropic dielectrics. The latter in particular suggests that enhanced effects should be observed for waves with slow group velocity. This result has in fact been recently demonstrated experimentally, where the relatively low group velocity of Alfv{\'e}n waves made it possible to control in laboratory experiments the rotation of the transverse structure of the wave by some tens of degrees over a few meters.

Accounting for and modeling the effects of motion on plasma waves, however, requires extending these findings to more general configurations than a uniform motion or a simple rotator. Considering this problem, it is then shown that applying Minkowski's theory to model propagation in media in non-uniform motion poses a number of challenges.  One of these challenges lies in that Minkowski's theory uses as input the rest-frame constitutive relations of a medium as input. Such formulation becomes problematic when considering non-uniform motion as one then needs to account for inertia. Although it is shown that workarounds may exist for specific motion, this remains an issue in general. An added difficulty is that the constitutive relations should be formulated in a way that does not affect the final wave equation. This is for instance not true of constitutive relations obtained assuming plane wave solutions. Indeed, it is demonstrated that because velocity non-uniformities manifest as inhomogeneities in the laboratory-frame, wave solutions cannot take the form of plane waves. Failing to recognize this result then leads to erroneous predictions, notably for polarization. However, it is shown that a consistent picture for the effect of motion, at least for the ray trajectory and the polarization, can be recovered from quasi-plane wave solutions when considering systematically how motion enters the wave equation and applying consistently Wigner–Weyl calculus. Although it has so far only been applied to isotropic media, this extended geometrical optics for moving media holds promise for plasmas.

To conclude, the scaling for drag effects obtained from simple models suggests that accounting for motion effects on plasma waves could be important, yet the models to do so in realistic configurations remain limited. Progress toward this goal, including the derivation of generic models, faces a number of basic challenges, which, if not addressed properly, can lead to errors. A recently explored yet promising pathway lies in the use of extended geometrical optics theory. Note finally that, to keep the discussion contained, we only focused here on the effect of motion on plasma waves, but that an equally important and compelling question is that of the effect of the wave on the plasma, and in particular how a wave can drive flows~\cite{Fetterman2008,Ochs2021a,Ochs2022,Rax2023a,Ochs2024}. Altogether, making progress on the interplay between waves and flows will advance our understanding of how waves can be used to diagnose or control plasmas, as desirable for instance in fusion applications~\cite{Rax2017,Kolmes2024}, but also of how flows may be used to manipulate waves.

\section*{Acknowledgments} 

This work was supported by the French Agence Nationale de la Recherche (ANR), under grant ANR-21-CE30-0002 (project WaRP). It has been carried out within the framework of the EUROfusion Consortium, via the Euratom Research and Training Programme (Grant Agreement No 101052200 – EUROfusion). Views and opinions expressed are however those of the authors only and do not necessarily reflect those of the European Union or the European Commission. 

\section*{Data availability statement} 

The data that supports the findings of this study are available within the article.

\section*{References} 
\bibliography{RefsAPSInvited2025}

@Article{Rax2023a,
  author  = {Rax, J.-M. and Gueroult, R. and Fisch, N. J.},
  journal = {J. Plasma. Phys.},
  title   = {Quasilinear theory of {B}rillouin resonances in rotating magnetized plasmas},
  year    = {2023},
  number  = {4},
  pages   = {905890408},
  volume  = {89},
  doi     = {10.1017/S0022377823000612},
}

@Article{Rax2023b,
  author    = {Rax, J.-M. and Gueroult, R. and Fisch, N.J.},
  journal   = {J. Plasma Phys.},
  title     = {Rotating {A}lfvén waves in rotating plasmas},
  year      = {2023},
  issn      = {1469-7807},
  month     = dec,
  number    = {6},
  pages     = {905890613},
  volume    = {89},
  doi       = {10.1017/s0022377823001368},
  publisher = {Cambridge University Press (CUP)},
}

@Article{Rax2021,
  author    = {J.-M. Rax and R. Gueroult},
  journal   = {J. Plasma Phys.},
  title     = {Faraday{\textendash}{F}resnel rotation and splitting of orbital angular momentum carrying waves in a rotating plasma},
  year      = {2021},
  month     = {sep},
  number    = {5},
  pages     = {905870507},
  volume    = {87},
  doi       = {10.1017/s0022377821000921},
  publisher = {Cambridge University Press ({CUP})},
}

@Article{Langlois2025,
  author    = {Langlois, J. and Braud, A. and Gueroult, R.},
  journal   = {J. Plasma Phys.},
  title     = {Fresnel drag in a moving magnetised plasma},
  year      = {2025},
  issn      = {1469-7807},
  month     = mar,
  number    = {2},
  pages     = {E47},
  volume    = {91},
  doi       = {10.1017/s0022377825000182},
  publisher = {Cambridge University Press (CUP)},
}

@Article{Langlois2024,
  author    = {Langlois, Julien and Gueroult, Renaud},
  journal   = {Proc. R. Soc. A.},
  title     = {Manifestations of inertia on light dragging revealed in plasmas},
  year      = {2024},
  issn      = {1471-2946},
  month     = nov,
  number    = {2301},
  pages     = {20240300},
  volume    = {480},
  doi       = {10.1098/rspa.2024.0300},
  publisher = {The Royal Society},
}

@Article{Langlois2023,
  author    = {Julien Langlois and Renaud Gueroult},
  journal   = {Phys. Rev. E},
  title     = {Contribution of fictitious forces to polarization drag in rotating media},
  year      = {2023},
  month     = {oct},
  number    = {4},
  pages     = {045201},
  volume    = {108},
  doi       = {10.1103/physreve.108.045201},
  publisher = {American Physical Society ({APS})},
}

@Article{Gueroult2024,
  author  = {Gueroult, R. and Tripathi, S.K.P. and Gaboriau, F. and Look, T.R. and Fisch, N.J.},
  journal = {J. Plasma Phys.},
  title   = {Plasma potential shaping using end-electrodes in the Large Plasma Device},
  year    = {2024},
  number  = {6},
  pages   = {905900603},
  volume  = {90},
  doi     = {10.1017/S0022377824000552},
}

@Article{Gueroult2019a,
  author   = {Gueroult, R. and Shi, Y. and Rax, J.-M. and Fisch, N. J.},
  journal  = {Nat. Commun.},
  title    = {Determining the rotation direction in pulsars},
  year     = {2019},
  month    = jul,
  number   = {1},
  pages    = {3232},
  volume   = {10},
  doi      = {10.1038/s41467-019-11243-4},
  refid    = {Gueroult2019},
}

@Article{Gueroult2020,
  author    = {Gueroult, R. and Rax, J.-M. and Fisch, N. J.},
  journal   = {Phys. Rev. E},
  title     = {Enhanced tuneable rotatory power in a rotating plasma},
  year      = {2020},
  month     = nov,
  number    = {5},
  pages     = {051202(R)},
  volume    = {102},
  doi       = {10.1103/PhysRevE.102.051202},
  publisher = {American Physical Society},
  refid     = {10.1103/PhysRevE.102.051202},
  url       = {http://dx.doi.org/10.1103/PhysRevE.102.051202},
}

@Article{Gueroult2023,
  author    = {Renaud Gueroult and Jean-Marcel Rax and Nathaniel J Fisch},
  journal   = {Plasma Phys. Control. Fusion},
  title     = {Wave propagation in rotating magnetised plasmas},
  year      = {2023},
  month     = {feb},
  number    = {3},
  pages     = {034006},
  volume    = {65},
  doi       = {10.1088/1361-6587/acb1d4},
  publisher = {{IOP} Publishing},
}

@Article{Gueroult2025,
  author    = {Gueroult, R. and Tripathi, S. K. and Han, J. and Pribyl, P. and Rax, J.-M. and Fisch, N. J.},
  journal   = {Phys. Rev. Lett.},
  title     = {Image rotation in plasmas},
  year      = {2025},
  issn      = {1079-7114},
  month     = may,
  pages     = {245101},
  volume    = {134},
  doi       = {10.1103/swrn-w3yf},
  publisher = {American Physical Society (APS)},
}

@Article{Gekelman2011,
  author    = {Gekelman, W. and Vincena, S. and Van Compernolle, B. and Morales, G. J. and Maggs, J. E. and Pribyl, P. and Carter, T. A.},
  journal   = {Phys. Plasmas},
  title     = {The many faces of shear Alfvén waves},
  year      = {2011},
  issn      = {1089-7674},
  month     = may,
  number    = {5},
  pages     = {055501},
  volume    = {18},
  doi       = {10.1063/1.3592210},
  publisher = {AIP Publishing},
}

@Article{Braud2025,
  author  = {Braud, A. and Langlois, J. and Gueroult, R.},
  journal = {Comptes Rendus. Physique},
  title   = {Geometrical optics methods for moving anisotropic media: a tool for magnetized plasmas},
  year    = {2025},
  pages   = {7},
  volume  = {26},
  doi     = {10.5802/crphys.218},
}

@Article{Lyutikov1998,
  author    = {Lyutikov, M.},
  journal   = {MNRAS},
  title     = {Waves in a one-dimensional magnetized relativistic pair plasma},
  year      = {1998},
  issn      = {1365-2966},
  month     = feb,
  number    = {4},
  pages     = {447--468},
  volume    = {293},
  doi       = {10.1046/j.1365-8711.1998.01154.x},
  publisher = {Oxford University Press (OUP)},
}

@Article{Rafat2019,
  author    = {Rafat, M. Z. and Melrose, D. B. and Mastrano, A.},
  journal   = {J. Plasma Phys.},
  title     = {Wave dispersion in pulsar plasma. Part 2. Pulsar frame},
  year      = {2019},
  issn      = {1469-7807},
  month     = jun,
  number    = {3},
 pages     = {905850311},
  volume    = {85},
  doi       = {10.1017/s0022377819000448},
  publisher = {Cambridge University Press (CUP)},
}

@Article{Fresnel1818,
  author  = {Fresnel, A.},
  journal = {Ann. Chim. Phys},
  title   = {Lettre d'{A}ugustin {F}resnel {\`a} {F}rançois {A}rago sur l' influence du mouvement terrestre dans quelques ph{\'e}nom{\`e}nes d'optique},
  year    = {1818},
  pages   = {57-66},
  volume  = {9},
  url     = {http://catalogue.bnf.fr/ark:/12148/cb343780820},
}

@Article{Fizeau1851,
  author  = {Fizeau, H.},
  journal = {C. R. Acad. Sci. Paris},
  title   = {Sur les hypoth{\`e}ses relatives {\`a} l'{\'e}ther lumineux, et sur une exp{\'e}rience qui para{\^i}t d{\'e}montrer que le mouvement des corps change la vitesse avec laquelle la lumi{\`e}re se propage dans leur int{\'e}rieur},
  year    = {1851},
  pages   = {349-355},
  volume  = {33},
  url     = {https://gallica.bnf.fr/ark:/12148/bpt6k29901/f351.image.r=Fizeau},
}

@Article{vonLaue1907,
  author  = {von {L}aue, M.},
  journal = {Ann. Phys.},
  title   = {Die Mitf{\"u}hrung des Lichtes durch bewegte K{\"o}rper nach dem Relativit{\"a}tsprinzip},
  year    = {1907},
  number  = {10},
  pages   = {989-990},
  volume  = {328},
  url     = {https://archive.org/details/70VonLaue},
}

@Article{Jones1976,
  author    = {R. V. Jones},
  journal   = {Proc. R. Soc. A},
  title     = {Rotary aether drag},
  year      = {1976},
  month     = {jun},
  number    = {1659},
  pages     = {423--439},
  volume    = {349},
  doi       = {10.1098/rspa.1976.0082},
  publisher = {The Royal Society},
  url       = {http://dx.doi.org/10.1098/rspa.1976.0082},
}

@Article{Jones1975,
  author    = {Jones, R. V.},
  journal   = {Proc. R. Soc. A},
  title     = {{\textquotesingle}{A}ether drag{\textquotesingle} in a transversely moving medium},
  year      = {1975},
  month     = {sep},
  number    = {1642},
  pages     = {351--364},
  volume    = {345},
  doi       = {10.1098/rspa.1975.0141},
  publisher = {The Royal Society},
}

@Article{Player1976,
  author  = {Player, M. A.},
  title   = {On the Dragging of the Plane of Polarization of Light Propagating in a Rotating Medium},
  journal = {Proc. R. Soc. A},
  year    = {1976},
  volume  = {349},
  number  = {1659},
  pages   = {441},
  month   = jun,
  doi     = {10.1098/rspa.1976.0083},
  url     = {http://dx.doi.org/10.1098/rspa.1976.0083},
}

@Article{Player1975,
  author    = {Player, M. A.},
  journal   = {Proc. R. Soc. A},
  title     = {Dispersion and the transverse aether drag},
  year      = {1975},
  month     = {sep},
  number    = {1642},
  pages     = {343--344},
  volume    = {345},
  doi       = {10.1098/rspa.1975.0139},
  publisher = {The Royal Society},
}

@Article{Goette2007,
  author   = {G{\"o}tte, J. B. and Barnett, S. M. and Padgett, M.},
  journal  = {Proc. R. Soc. A},
  title    = {On the dragging of light by a rotating medium},
  year     = {2007},
  month    = sep,
  number   = {2085},
  pages    = {2185},
  volume   = {463},
  doi      = {10.1098/rspa.2007.1871},
  url      = {http://dx.doi.org/10.1098/rspa.2007.1871},
}

@Article{Franke-Arnold2011,
  author   = {Franke-Arnold, S. and Gibson, G. and Boyd, R. W. and Padgett, M. J.},
  journal  = {Science},
  title    = {Rotary Photon Drag Enhanced by a Slow-Light Medium},
  year     = {2011},
  month    = jun,
  number   = {6038},
  pages    = {65},
  volume   = {333},
  doi      = {10.1126/science.1203984},
  url      = {http://dx.doi.org/10.1126/science.1203984},
}

@article{Artoni2001,
  title={Fresnel light drag in a coherently driven moving medium},
  author={Artoni, Maurizio and Carusotto, Iacopo and La Rocca, Giuseppe Carlo and Bassani, F},
  journal={Phys. Rev. Lett.},
  volume={86},
  number={12},
  pages={2549},
  year={2001},
  publisher={APS},
  doi={10.1103/PhysRevLett.86.2549}
}

@article{Safari2016,
  title={Light-drag enhancement by a highly dispersive rubidium vapor},
  author={Safari, Akbar and De Leon, Israel and Mirhosseini, Mohammad and Maga{\~n}a-Loaiza, Omar S and Boyd, Robert W},
  journal={Phys. Rev. Lett.},
  volume={116},
  number={1},
  pages={013601},
  year={2016},
  publisher={APS},
  doi={10.1103/PhysRevLett.116.013601}
}

@book{Landau2013,
  title={Electrodynamics of continuous media},
  author={Landau, Lev Davidovich and Bell, John Stewart and Kearsley, MJ and Pitaevskii, LP and Lifshitz, EM and Sykes, JB},
  volume={8},
  year={2013},
  publisher={Elsevier}
}

@Book{Sommerfeld1952,
  author    = {Sommerfeld, Arnold},
  publisher = {Academic Press, New York},
  title     = {Electrodynamics: lectures on theoretical physics, vol {III}},
  year      = {1952},
}

@book{Whittaker1989,
  title={A History of the Theories of Aether and Electricity: Vol. I: The Classical Theories; Vol. II: The Modern Theories, 1900-1926},
  author={Whittaker, Edmund},
  volume={1},
  year={1989},
  publisher={Courier Dover Publications}
}

@Article{Steinitz2020,
  author    = {U. Steinitz and I. S. Averbukh},
  journal   = {Phys. Rev. A},
  title     = {Giant polarization drag in a gas of molecular super-rotors},
  year      = {2020},
  month     = {feb},
  number    = {2},
  pages     = {021404},
  volume    = {101},
  doi       = {10.1103/physreva.101.021404},
  publisher = {American Physical Society ({APS})},
}

@Article{Milner2021,
  author    = {A. A. Milner and U. Steinitz and I. S. Averbukh and V. Milner},
  journal   = {Phys. Rev. Lett.},
  title     = {Observation of Mechanical {F}araday Effect in Gaseous Media},
  year      = {2021},
  month     = {aug},
  number    = {7},
  pages     = {073901},
  volume    = {127},
  doi       = {10.1103/physrevlett.127.073901},
  publisher = {American Physical Society ({APS})},
}

@Article{Fermi1923,
  author  = {Fermi, E.},
  journal = {Rend. Mat. Acc. Lincei},
  title   = {Sul trascinamento del piano di polarizzazione da parte di un messo rotante},
  year    = {1923},
  note    = {Reprinted in Collected Papers, vol. 1 (University of Chicago Press, Chicago, 1962)},
  pages   = {115--118},
  volume  = {32},
}

@Article{Rice2007,
  author    = {Rice, J.E. and Ince-Cushman, A. and deGrassie, J.S. and Eriksson, L.-G. and Sakamoto, Y. and Scarabosio, A. and Bortolon, A. and Burrell, K.H. and Duval, B.P. and Fenzi-Bonizec, C. and Greenwald, M.J. and Groebner, R.J. and Hoang, G.T. and Koide, Y. and Marmar, E.S. and Pochelon, A. and Podpaly, Y.},
  journal   = {Nucl. Fusion},
  title     = {Inter-machine comparison of intrinsic toroidal rotation in tokamaks},
  year      = {2007},
  issn      = {1741-4326},
  month     = oct,
  number    = {11},
  pages     = {1618--1624},
  volume    = {47},
  doi       = {10.1088/0029-5515/47/11/025},
  publisher = {IOP Publishing},
}

@InBook{Komissarov2012,
  author    = {Komissarov, Serguei},
  chapter   = {Central Engines: Acceleration, Collimation and Confinement of Jets},
  pages     = {81--114},
  publisher = {Wiley},
  title     = {Relativistic Jets from Active Galactic Nuclei},
  year      = {2012},
  isbn      = {9783527641741},
  month     = jan,
  doi       = {10.1002/9783527641741.ch4},
}

@Article{Mirabel2003,
  author    = {Mirabel, I.F.},
  journal   = {New Astron. Rev.},
  title     = {Relativistic jets in the universe},
  year      = {2003},
  issn      = {1387-6473},
  month     = oct,
  number    = {6–7},
  pages     = {471--475},
  volume    = {47},
  doi       = {10.1016/s1387-6473(03)00074-5},
  publisher = {Elsevier BV},
}

@Article{Bailey1948,
  author    = {Bailey, VA},
  journal   = {Austrl. J. Sci. Res. A},
  title     = {Plane Waves in an Ionized Gas with Static Electric and Magnetic Fields Present},
  year      = {1948},
  issn      = {0365-3676},
  month     = dec,
  number    = {4},
  pages     = {351--359},
  volume    = {1},
  doi       = {10.1071/ch9480351},
  publisher = {CSIRO Publishing},
}

@Article{Lehnert1954,
  author    = {Lehnert, B.},
  journal   = {ApJ},
  title     = {Magnetohydrodynamic Waves Under the Action of the Coriolis Force.},
  year      = {1954},
  issn      = {1538-4357},
  month     = may,
  pages     = {647},
  volume    = {119},
  doi       = {10.1086/145869},
  publisher = {American Astronomical Society},
}

@Article{Lehnert1955,
  author    = {Lehnert, B.},
  journal   = {ApJ},
  title     = {Magnetohydrodynamic Waves Under the Action of the Coriolis Force. II.},
  year      = {1955},
  issn      = {1538-4357},
  month     = mar,
  pages     = {481},
  volume    = {121},
  doi       = {10.1086/146009},
  publisher = {American Astronomical Society},
}

@Article{Tandon1966,
  author    = {Tandon, J. N. and Bajaj, N. K.},
  journal   = {MNRAS},
  title     = {Wave Propagation in Rarefied Plasma Under the Action of the Coriolis Force},
  year      = {1966},
  issn      = {0035-8711},
  month     = apr,
  number    = {3},
  pages     = {285--304},
  volume    = {132},
  doi       = {10.1093/mnras/132.2.285},
  publisher = {Oxford University Press (OUP)},
}

@Article{Uberoi1970,
  author    = {Uberoi, C and Das, G C},
  journal   = {Plasma Physics},
  title     = {Wave propagation in cold plasma in the presence of the Coriolis force},
  year      = {1970},
  issn      = {0032-1028},
  month     = sep,
  number    = {9},
  pages     = {661--684},
  volume    = {12},
  doi       = {10.1088/0032-1028/12/9/002},
  publisher = {IOP Publishing},
}

@Article{Verheest1974,
  author    = {Verheest, Frank},
  journal   = {Astrophy. Space Sci.},
  title     = {Dispersion and stability of waves in plasmas in the presence of a coriolis force},
  year      = {1974},
  issn      = {1572-946X},
  month     = may,
  number    = {1},
  pages     = {91--99},
  volume    = {28},
  doi       = {10.1007/bf00642239},
  publisher = {Springer Science and Business Media LLC},
}

@Article{Engels1975,
  author    = {Engels, E. and Verheest, F.},
  journal   = {Astrophy. Space Sci.},
  title     = {Wave propagation in rotating plasmas and influence of the Coriolis force},
  year      = {1975},
  issn      = {1572-946X},
  month     = oct,
  number    = {2},
  pages     = {427--440},
  volume    = {37},
  doi       = {10.1007/bf00640362},
  publisher = {Springer Science and Business Media LLC},
}

@Article{Minkowski1908,
  author  = {Minkowski, H.},
  journal = {Nachrichten von der Gesellschaft der Wissenschaften zu Göttingen, Mathematisch-Physikalische Klasse},
  title   = {Die Grundgleichungen für die elektromagnetischen Vorgänge in bewegten Körpern},
  year    = {1908},
  pages   = {53-111},
  volume  = {1908},
  url     = {http://eudml.org/doc/58707},
}

@Article{Tai1965,
  author    = {Tai, C.T.},
  journal   = {J. Res. Natl. Bur. Std.},
  title     = {Electrodynamics of moving anisotropic media: The First-order theory},
  year      = {1965},
  issn      = {0502-2568},
  month     = mar,
  number    = {3},
  pages     = {401},
  volume    = {69D},
  doi       = {10.6028/jres.069d.051},
  publisher = {National Institute of Standards and Technology (NIST)},
}

@Article{Melrose1973,
  author    = {Melrose, D B},
  journal   = {Plasma Physics},
  title     = {A covariant formulation of wave dispersion},
  year      = {1973},
  issn      = {0032-1028},
  month     = feb,
  number    = {2},
  pages     = {99--106},
  volume    = {15},
  doi       = {10.1088/0032-1028/15/2/002},
  publisher = {IOP Publishing},
}

@Article{Hebenstreit1979,
  author    = {Hebenstreit, Helmut},
  journal   = {Z. fur Naturforsch. A},
  title     = {Calculation of Covariant Dispersion Equations for Moving Plasmas},
  year      = {1979},
  issn      = {0932-0784},
  month     = feb,
  number    = {2},
  pages     = {155--162},
  volume    = {34},
  doi       = {10.1515/zna-1979-0205},
  publisher = {Walter de Gruyter GmbH},
}

@Article{Lee1966,
  author    = {Lee, S. W. and Lo, Y. T.},
  journal   = {Radio Sci.},
  title     = {Radiation in a Moving Anisotropic Medium},
  year      = {1966},
  issn      = {1944-799X},
  month     = mar,
  number    = {3},
  pages     = {313--324},
  volume    = {1},
  doi       = {10.1002/rds196613313},
  publisher = {American Geophysical Union (AGU)},
}

@Article{Chen1966,
  author    = {Chen, H.C. and Cheng, D.K.},
  journal   = {Proc. IEEE},
  title     = {Constitutive relations for a moving anisotropic medium},
  year      = {1966},
  issn      = {0018-9219},
  number    = {1},
  pages     = {62--63},
  volume    = {54},
  doi       = {10.1109/proc.1966.4581},
  publisher = {Institute of Electrical and Electronics Engineers (IEEE)},
}

@Article{VanBladel1976,
  author    = {Van Bladel, J.},
  journal   = {Proc. IEEE},
  title     = {Electromagnetic fields in the presence of rotating bodies},
  year      = {1976},
  issn      = {0018-9219},
  number    = {3},
  pages     = {301--318},
  volume    = {64},
  doi       = {10.1109/proc.1976.10111},
  publisher = {Institute of Electrical and Electronics Engineers (IEEE)},
}

@Article{Shiozawa1973,
  author    = {Shiozawa, T.},
  journal   = {Proc. IEEE},
  title     = {Phenomenological and electron-theoretical study of the electrodynamics of rotating systems},
  year      = {1973},
  issn      = {0018-9219},
  number    = {12},
  pages     = {1694--1702},
  volume    = {61},
  doi       = {10.1109/proc.1973.9359},
  publisher = {Institute of Electrical and Electronics Engineers (IEEE)},
}

@Article{Baranova1979,
  author    = {Baranova, N. B. and Zel'dovich, Y. B.},
  journal   = {Proc. R. Soc. A.},
  title     = {Coriolis contribution to the rotatory ether drag},
  year      = {1979},
  issn      = {0080-4630},
  month     = nov,
  number    = {1735},
  pages     = {591--592},
  volume    = {368},
  doi       = {10.1098/rspa.1979.0148},
  publisher = {The Royal Society},
}

@Article{Unz1966,
  author    = {Unz, H.},
  journal   = {Radio Sci.},
  title     = {Wave Propagation in Drifting Isotropic Warm Plasma},
  year      = {1966},
  issn      = {1944-799X},
  month     = mar,
  number    = {3},
  pages     = {325--338},
  volume    = {1},
  doi       = {10.1002/rds196613325},
  publisher = {American Geophysical Union (AGU)},
}

@Article{Chawla1966,
  author    = {Chawla, B. R. and Rao, S. S. and Unz, H.},
  journal   = {J. Appl. Phys.},
  title     = {Wave Propagation in Homogeneous Gyrotropic Compressible Drifting Plasma},
  year      = {1966},
  issn      = {1089-7550},
  month     = aug,
  number    = {9},
  pages     = {3563--3566},
  volume    = {37},
  doi       = {10.1063/1.1708902},
  publisher = {AIP Publishing},
}

@Article{Unz1962,
  author    = {Unz, H.},
  journal   = {IRE Trans. Anntenas Propag.},
  title     = {The magneto-ionic theory for drifting plasma},
  year      = {1962},
  issn      = {0096-1973},
  month     = jul,
  number    = {4},
  pages     = {459--464},
  volume    = {10},
  doi       = {10.1109/tap.1962.1137893},
  publisher = {Institute of Electrical and Electronics Engineers (IEEE)},
}

@Article{Talekar1974,
  author    = {Talekar, V L and Bhandari, S S},
  journal   = {Plasma Physics},
  title     = {Wave propagation in drifting weak magnetoplasma under {G}rad’s thirteen-moment approximation},
  year      = {1974},
  issn      = {0032-1028},
  month     = aug,
  number    = {8},
  pages     = {741--751},
  volume    = {16},
  doi       = {10.1088/0032-1028/16/8/004},
  publisher = {IOP Publishing},
}

@Article{Chandrasekhar1953,
  author    = {S. Chandrasekhar},
  journal   = {MNRAS},
  title     = {Problems of Stability in Hydrodynamics and Hydromagnetics: {G}eorge {D}arwin Lecture, delivered by Professor {S}. {C}handrasekhar on 1953 Novemher 13},
  year      = {1953},
  month     = {dec},
  number    = {6},
  pages     = {667--678},
  volume    = {113},
  doi       = {10.1093/mnras/113.6.667},
  publisher = {Oxford University Press ({OUP})},
}

@Article{Shiozawa1974,
  author    = {Shiozawa, T.},
  journal   = {Proc. IEEE},
  title     = {Constitutive relations for a rotating electron plasma},
  year      = {1974},
  issn      = {0018-9219},
  number    = {9},
  pages     = {1283--1284},
  volume    = {62},
  doi       = {10.1109/proc.1974.9611},
  publisher = {Institute of Electrical and Electronics Engineers (IEEE)},
}

@Article{Scarf1961,
  author    = {Scarf, F. L.},
  journal   = {Am. J. Phys.},
  title     = {Wave Propagation in a Moving Plasma},
  year      = {1961},
  issn      = {1943-2909},
  month     = feb,
  number    = {2},
  pages     = {101--107},
  volume    = {29},
  doi       = {10.1119/1.1937681},
  publisher = {American Association of Physics Teachers (AAPT)},
}

@Article{Chawla1967,
  author    = {Chawla, B.R. and Unz, H.},
  journal   = {Electron. Lett.},
  title     = {Wave propagation in a moving plasma},
  year      = {1967},
  issn      = {1350-911X},
  month     = jun,
  number    = {6},
  pages     = {244--246},
  volume    = {3},
  doi       = {10.1049/el:19670189},
  publisher = {Institution of Engineering and Technology (IET)},
}

@Article{Unz1966a,
  author    = {Unz, H.},
  journal   = {Phys. Rev.},
  title     = {Relativistic Magneto-Ionic Theory for Drifting Plasma in Longitudinal Direction},
  year      = {1966},
  issn      = {0031-899X},
  month     = jun,
  number    = {1},
  pages     = {92--95},
  volume    = {146},
  doi       = {10.1103/physrev.146.92},
  publisher = {American Physical Society (APS)},
}

@Article{Unz1968,
  author    = {Unz, H.},
  journal   = {Il Nuovo Cimento B},
  title     = {Radiation from electromagnetic sources in an unbounded warm magneto-plasma},
  year      = {1968},
  issn      = {1826-9877},
  month     = jun,
  number    = {2},
  pages     = {481--490},
  volume    = {55},
  doi       = {10.1007/bf02711657},
  publisher = {Springer Science and Business Media LLC},
}

@Article{Censor2010,
  author    = {Censor, D.},
  journal   = {ZAMM},
  title     = {Relativistic invariance of dispersion‐relations and their associated wave‐operators and Green‐functions},
  year      = {2010},
  issn      = {1521-4001},
  month     = feb,
  number    = {3},
  pages     = {194--202},
  volume    = {90},
  doi       = {10.1002/zamm.200900298},
  publisher = {Wiley},
}

@Article{Censor1980,
  author    = {Censor, D.},
  journal   = {Proc. IEEE},
  title     = {Dispersion equations in moving media},
  year      = {1980},
  issn      = {0018-9219},
  number    = {4},
  pages     = {528--529},
  volume    = {68},
  doi       = {10.1109/proc.1980.11677},
  publisher = {Institute of Electrical and Electronics Engineers (IEEE)},
}

@Book{Kong2008,
  author    = {Kong, J. A.},
  publisher = {EMW Publishing},
  title     = {Electromagnetic wave theory},
  year      = {2008},
}

@Article{Thomson1885,
  author  = {Thomson, J. J.},
  journal = {Proc. Camb. Phil. Soc.},
  title   = {Note on the Rotation of the Plane of Polarization of Light by a Moving Medium},
  year    = {1885},
  pages   = {250},
  volume  = {5},
  url     = {https://www.biodiversitylibrary.org/page/26955928},
}

@Article{Faraday1846,
  author  = {Faraday, M.},
  title   = {On the Magnetization of Light and the Illumination of Magnetic Lines of Force},
  journal = {Phil. Trans. Roy. Soc. London},
  year    = {1846},
  volume  = {136},
  number  = {1},
  pages   = {1-20},
  doi     = {10.1098/rstl.1846.0001},
  url     = {http://dx.doi.org/10.1098/rstl.1846.0001},
}

@Article{Carusotto2003,
  author    = {Carusotto, I. and Artoni, M. and La Rocca, G. C. and Bassani, F.},
  journal   = {Phys. Rev. A},
  title     = {Transverse {F}resnel-{F}izeau drag effects in strongly dispersive media},
  year      = {2003},
  issn      = {1094-1622},
  month     = dec,
  number    = {6},
  pages     = {063819},
  volume    = {68},
  doi       = {10.1103/physreva.68.063819},
  publisher = {American Physical Society (APS)},
}

@Article{Braud2025a,
  author    = {Braud, Aymeric and Gueroult, Renaud},
  journal   = {Phys. Rev. A},
  title     = {Spin-orbit interactions induced by light drag in moving media},
  year      = {2025},
  issn      = {2469-9934},
  month     = oct,
  number    = {4},
  pages     = {043505},
  volume    = {112},
  doi       = {10.1103/lsgb-ylzf},
  publisher = {American Physical Society (APS)},
}

@Article{Arnaud1976,
  author    = {Arnaud, J. A.},
  journal   = {Nature},
  title     = {Dispersion and the transverse aether drag},
  year      = {1976},
  issn      = {1476-4687},
  month     = jun,
  number    = {5560},
  pages     = {481--482},
  volume    = {261},
  doi       = {10.1038/261481a0},
  publisher = {Springer Science and Business Media LLC},
}

@Article{Ko1978,
  author    = {Ko, H. C. and Chuang, C. W.},
  journal   = {ApJ},
  title     = {On the passage of radiation through moving astrophysical plasmas},
  year      = {1978},
  issn      = {1538-4357},
  month     = jun,
  pages     = {1012},
  volume    = {222},
  doi       = {10.1086/156219},
  publisher = {American Astronomical Society},
}

@Article{MeyerVernet1980,
  author    = {Meyer-Vernet, N.},
  journal   = {Astrophy. Space Sci.},
  title     = {High-frequency transverse fresnel drag in a moving magneto-active plasma},
  year      = {1980},
  issn      = {1572-946X},
  month     = nov,
  number    = {1},
  pages     = {207--212},
  volume    = {73},
  doi       = {10.1007/bf00642376},
  publisher = {Springer Science and Business Media LLC},
}

@Article{Mukherjee1975,
  author    = {Mukherjee, P. K.},
  journal   = {J. Appl. Phys.},
  title     = {Electromagnetic wave propagation in a moving magnetoplasma medium in the presence of a boundary},
  year      = {1975},
  issn      = {1089-7550},
  month     = may,
  number    = {5},
  pages     = {2295--2297},
  volume    = {46},
  doi       = {10.1063/1.321825},
  publisher = {AIP Publishing},
}

@Article{Padgett2006,
  author    = {Padgett, Miles and Whyte, Graeme and Girkin, John and Wright, Amanda and Allen, Les and Öhberg, Patrik and Barnett, Stephen M.},
  journal   = {Opt. Lett.},
  title     = {Polarization and image rotation induced by a rotating dielectric rod: an optical angular momentum interpretation},
  year      = {2006},
  issn      = {1539-4794},
  month     = jul,
  number    = {14},
  pages     = {2205},
  volume    = {31},
  doi       = {10.1364/ol.31.002205},
  publisher = {Optica Publishing Group},
}

@Article{Mendonca2009,
  author    = {Mendonca, J. T. and Ali, S. and Thidé, B.},
  journal   = {Phys. Plasmas},
  title     = {Plasmons with orbital angular momentum},
  year      = {2009},
  issn      = {1089-7674},
  month     = nov,
  number    = {11},
  pages     = {112103},
  volume    = {16},
  doi       = {10.1063/1.3261802},
  publisher = {AIP Publishing},
}

@Article{Mendonca2012,
  author    = {Mendonça, J T},
  journal   = {Plasma Phys. Contr. F.},
  title     = {Twisted waves in a plasma},
  year      = {2012},
  issn      = {1361-6587},
  month     = nov,
  number    = {12},
  pages     = {124031},
  volume    = {54},
  doi       = {10.1088/0741-3335/54/12/124031},
  publisher = {IOP Publishing},
}

@Article{Mendonca2012a,
  author    = {Mendonça, J. T.},
  journal   = {Phys. Plasmas},
  title     = {Kinetic description of electron plasma waves with orbital angular momentum},
  year      = {2012},
  issn      = {1089-7674},
  month     = nov,
  number    = {11},
  volume    = {19},
  pages     = {112113},
  doi       = {10.1063/1.4769030},
  publisher = {AIP Publishing},
}

@Article{Chen2017,
  author    = {Chen, Qiang and Qin, Hong and Liu, Jian},
  journal   = {Sci. Rep.},
  title     = {Photons, phonons, and plasmons with orbital angular momentum in plasmas},
  year      = {2017},
  issn      = {2045-2322},
  month     = feb,
  number    = {1},
  volume    = {7},
  pages     = {41731},
  doi       = {10.1038/srep41731},
  publisher = {Springer Science and Business Media LLC},
}

@Article{Urrutia2016,
  author    = {Urrutia, J. M. and Stenzel, R. L.},
  journal   = {Phys. Plasmas},
  title     = {Helicon waves in uniform plasmas. IV. Bessel beams, Gendrin beams, and helicons},
  year      = {2016},
  issn      = {1089-7674},
  month     = may,
  number    = {5},
  pages     = {052112},
  volume    = {23},
  doi       = {10.1063/1.4949348},
  publisher = {AIP Publishing},
}

@Article{Stenzel2015,
  author    = {Stenzel, R. L. and Urrutia, J. M.},
  journal   = {Phys. Plasmas},
  title     = {Helicon waves in uniform plasmas. II. High m numbers},
  year      = {2015},
  issn      = {1089-7674},
  month     = sep,
  number    = {9},
  pages     = {092113},
  volume    = {22},
  doi       = {10.1063/1.4930106},
  publisher = {AIP Publishing},
}

@Article{Stenzel2016,
  author    = {Stenzel, R. L.},
  journal   = {Adv. Phys. X},
  title     = {Whistler waves with angular momentum in space and laboratory plasmas and their counterparts in free space},
  year      = {2016},
  issn      = {2374-6149},
  month     = jul,
  number    = {4},
  pages     = {687--710},
  volume    = {1},
  doi       = {10.1080/23746149.2016.1240017},
  publisher = {Informa UK Limited},
}

@Article{Shukla2012,
  author    = {Shukla, P.K.},
  journal   = {Phys. Lett. A},
  title     = {Twisted shear Alfvén waves with orbital angular momentum},
  year      = {2012},
  issn      = {0375-9601},
  month     = sep,
  number    = {44},
  pages     = {2792--2794},
  volume    = {376},
  doi       = {10.1016/j.physleta.2012.08.025},
  publisher = {Elsevier BV},
}

@Article{Gekelman2016,
  author   = {Gekelman, W. and Pribyl, P. and Lucky, Z. and Drandell, M. and Leneman, D. and Maggs, J. and Vincena, S. and Compernolle, B. Van and Tripathi, S. K. P. and Morales, G. and Carter, T. A. and Wang, Y. and DeHaas, T.},
  title    = {The upgraded Large Plasma Device, a machine for studying frontier basic plasma physics},
  journal  = {Rev. Sci. Instrum.},
  year     = {2016},
  volume   = {87},
  number   = {2},
  pages    = {025105},
  doi      = {10.1063/1.4941079},
  url      = {http://dx.doi.org/10.1063/1.4941079},
}

@Article{WisniewskiBarker2014,
  author    = {Wisniewski-Barker, Emma and Gibson, Graham M. and Franke-Arnold, Sonja and Boyd, Robert W. and Padgett, Miles J.},
  journal   = {Opt. Express},
  title     = {Mechanical Faraday effect for orbital angular momentum-carrying beams},
  year      = {2014},
  issn      = {1094-4087},
  month     = may,
  number    = {10},
  pages     = {11690},
  volume    = {22},
  doi       = {10.1364/oe.22.011690},
  publisher = {Optica Publishing Group},
}

@Book{Ginzburg1964,
  author    = {Ginzburg, V. L.},
  publisher = {Addison-Wesley, Reading,Mass.},
  title     = {The Propagation Of Electromagnetic Waves In Plasmas},
  year      = {1964},
}

@Book{Misner2017,
  author    = {Misner, C. W. and Thorne, K. S. and Wheeler, J. A.},
  publisher = {Princeton University Press},
  title     = {Gravitation},
  year      = {2017},
}

@Book{Tracy2014,
  author    = {Tracy, E. R. and Brizard, A. J. and Richardson, A. S. and Kaufman, A. N.},
  publisher = {Cambridge University Press},
  title     = {Ray Tracing and Beyond: Phase Space Methods in Plasma Wave Theory},
  year      = {2014},
  isbn      = {9780511667565},
  month     = feb,
  doi       = {10.1017/cbo9780511667565},
}

@Article{Ruiz2017,
  author    = {Ruiz, D. E. and Dodin, I. Y.},
  journal   = {Phys. Plasmas},
  title     = {Extending geometrical optics: A {L}agrangian theory for vector waves},
  year      = {2017},
  issn      = {1089-7674},
  month     = mar,
  number    = {5},
  volume    = {24},
  pages     = {055704},
  doi       = {10.1063/1.4977537},
  publisher = {AIP Publishing},
}

@Article{Dodin2019,
  author    = {Dodin, I. Y. and Ruiz, D. E. and Yanagihara, K. and Zhou, Y. and Kubo, S.},
  journal   = {Phys. Plasmas},
  title     = {Quasioptical modeling of wave beams with and without mode conversion. I. Basic theory},
  year      = {2019},
  issn      = {1089-7674},
  month     = jul,
  number    = {7},
  volume    = {26},
  doi       = {10.1063/1.5095076},
  publisher = {AIP Publishing},
}

@PhdThesis{Ruiz2017a,
  author = {Ruiz, Daniel Edward},
  school = {Princeton University},
  title  = {A geometric theory of waves and its applications to plasma physics},
  year   = {2017},
  url    = {http://arks.princeton.edu/ark:/88435/dsp01k930c068w},
}

@Article{Ruiz2015,
  author    = {Ruiz, D. E. and Dodin, I. Y.},
  journal   = {Phys. Rev. A},
  title     = {First-principles variational formulation of polarization effects in geometrical optics},
  year      = {2015},
  issn      = {1094-1622},
  month     = oct,
  number    = {4},
  pages     = {043805},
  volume    = {92},
  doi       = {10.1103/physreva.92.043805},
  publisher = {American Physical Society (APS)},
}

@Book{Kravtsov1990,
  author    = {Kravtsov, Yu. A. and Orlov, Yu. I.},
  publisher = {Springer-Verlag, New York},
  title     = {Geometrical opticsof inhomogeneous media},
  year      = {1990},
}

@Article{Bliokh2015,
  author    = {Bliokh, K. Y. and Rodríguez-Fortuño, F. J. and Nori, F. and Zayats, A. V.},
  journal   = {Nat. Photonics},
  title     = {Spin–orbit interactions of light},
  year      = {2015},
  issn      = {1749-4893},
  month     = nov,
  number    = {12},
  pages     = {796--808},
  volume    = {9},
  doi       = {10.1038/nphoton.2015.201},
  publisher = {Springer Science and Business Media LLC},
}

@Article{Braud2026,
  author    = {Braud, Aymeric and Gueroult, Renaud},
  journal   = {Phys. Rev. A},
  title     = {Spin-redirection Berry phase with planar rays},
  year      = {2026},
  month     = {May},
  pages     = {L051501},
  volume    = {113},
  doi       = {10.1103/vkls-qpkg},
  issue     = {5},
  numpages  = {7},
  publisher = {American Physical Society},
}

@Book{Born2019,
  author    = {Born, Max and Wolf, Emil},
  publisher = {Cambridge University Press},
  title     = {Principles of Optics: 60th Anniversary Edition},
  year      = {2019},
  isbn      = {9781108477437},
  month     = dec,
  doi       = {10.1017/9781108769914},
}

@Article{Anderson1969,
  author    = {Anderson, J. L. and Ryon, J. W.},
  journal   = {Phys. Rev.},
  title     = {Electromagnetic Radiation in Accelerated Systems},
  year      = {1969},
  issn      = {0031-899X},
  month     = may,
  number    = {5},
  pages     = {1765--1775},
  volume    = {181},
  doi       = {10.1103/physrev.181.1765},
  publisher = {American Physical Society (APS)},
}

@Article{Steinberg2023,
  author    = {Steinberg, Ben Z. and Engheta, Nader},
  journal   = {Phys. Rev. B},
  title     = {Rest-frame quasistatic theory for rotating electromagnetic systems and circuits},
  year      = {2023},
  issn      = {2469-9969},
  month     = may,
  number    = {19},
  pages     = {195418},
  volume    = {107},
  doi       = {10.1103/physrevb.107.195418},
  publisher = {American Physical Society (APS)},
}

@Article{Ochs2021a,
  author    = {Ochs, Ian E. and Fisch, Nathaniel J.},
  journal   = {Phys. Rev. Lett.},
  title     = {Nonresonant Diffusion in Alpha Channeling},
  year      = {2021},
  issn      = {1079-7114},
  month     = jul,
  number    = {2},
  pages     = {025003},
  volume    = {127},
  doi       = {10.1103/physrevlett.127.025003},
  publisher = {American Physical Society (APS)},
}

@Article{Ochs2024,
  author    = {Ochs, Ian E.},
  journal   = {Phys. Plasmas},
  title     = {When do waves drive plasma flows?},
  year      = {2024},
  issn      = {1089-7674},
  month     = apr,
  number    = {4},
  volume    = {31},
  pages     = {042116},
  doi       = {10.1063/5.0201780},
  publisher = {AIP Publishing},
}

@Article{Fetterman2008,
  author    = {Fetterman, Abraham J. and Fisch, Nathaniel J.},
  journal   = {Phys. Rev. Lett.},
  title     = {$\alpha$ Channeling in a Rotating Plasma},
  year      = {2008},
  issn      = {1079-7114},
  month     = nov,
  number    = {20},
  pages     = {205003},
  volume    = {101},
  doi       = {10.1103/physrevlett.101.205003},
  publisher = {American Physical Society (APS)},
}

@Article{Rax2017,
  author    = {Rax, J. M. and Gueroult, R. and Fisch, N. J.},
  journal   = {Phys. Plasmas},
  title     = {Efficiency of wave-driven rigid body rotation toroidal confinement},
  year      = {2017},
  issn      = {1089-7674},
  month     = mar,
  number    = {3},
  volume    = {24},
  pages     = {032504},
  doi       = {10.1063/1.4977919},
  publisher = {AIP Publishing},
}

@Article{Kolmes2024,
  author    = {Kolmes, E. J. and Ochs, I. E. and Rax, J.-M. and Fisch, N. J.},
  journal   = {Nat. Commun.},
  title     = {Massive, long-lived electrostatic potentials in a rotating mirror plasma},
  year      = {2024},
  issn      = {2041-1723},
  month     = may,
  number    = {1},
  volume    = {15},
  pages     = {4302},
  doi       = {10.1038/s41467-024-47386-2},
  publisher = {Springer Science and Business Media LLC},
}

@Article{Ochs2022,
  author    = {Ochs, Ian E. and Fisch, Nathaniel J.},
  journal   = {Phys. Plasmas},
  title     = {Momentum conservation in current drive and alpha-channeling-mediated rotation drive},
  year      = {2022},
  issn      = {1089-7674},
  month     = jun,
  number    = {6},
  volume    = {29},
  pages     = {062106},
  doi       = {10.1063/5.0085821},
  publisher = {AIP Publishing},
}

@Article{Rozanov2005,
  author    = {Rozanov, N. N. and Sochilin, G. B.},
  journal   = {Opt. Spectrosc.},
  title     = {Geometrical optics of moving media},
  year      = {2005},
  issn      = {1562-6911},
  month     = mar,
  number    = {3},
  pages     = {441--446},
  volume    = {98},
  doi       = {10.1134/1.1890525},
  publisher = {Pleiades Publishing Ltd},
}

@Article{Rozanov2006,
  author    = {Rozanov, Nikolai N. and Sochilin, G.B.},
  journal   = {Usp. Fiz. Nauk.},
  title     = {First order relativistic effects in the electrodynamics of media moving with a nonuniform velocity},
  year      = {2006},
  issn      = {1996-6652},
  number    = {4},
  pages     = {421},
  volume    = {176},
  doi       = {10.3367/ufnr.0176.200604f.0421},
  publisher = {Uspekhi Fizicheskikh Nauk (UFN) Journal},
}

@Article{Langlois2025a,
  title={Light drag in nonuniformly moving anisotropic media through the lens of gradient-index optics},
  author={Langlois, Julien and Gueroult, Renaud},
  journal={Phys. Rev. E},
  volume={113},
  number={4},
  pages={045205},
  year={2026},
  publisher={APS},
  doi={10.1103/8gx6-rnrf}
}

@Article{Banerjee2022,
  author    = {Banerjee, Chitram and Solomons, Yakov and Black, A. Nicholas and Marcucci, Giulia and Eger, David and Davidson, Nir and Firstenberg, Ofer and Boyd, Robert W.},
  journal   = {Phys. Rev. Research},
  title     = {Anomalous optical drag},
  year      = {2022},
  issn      = {2643-1564},
  month     = aug,
  number    = {3},
  pages     = {033124},
  volume    = {4},
  doi       = {10.1103/physrevresearch.4.033124},
  publisher = {American Physical Society (APS)},
}

@Article{Jones1972,
  author    = {Jones, R. V.},
  journal   = {Proc. R. Soc. A},
  title     = {{\textquotesingle}{F}resnel aether drag{\textquotesingle} in a transversely moving medium},
  year      = {1972},
  month     = {jun},
  number    = {1574},
  pages     = {337--352},
  volume    = {328},
  doi       = {10.1098/rspa.1972.0081},
  publisher = {The Royal Society},
}

\appendix

\section{Wave equation for $\mathbf{E}$ for a rotating dielectric from Minkowski's theory}
\label{App:WaveE}

Let us show here how one may derive a wave equation for $\mathbf{E}$ from  Minkoswki's theory. Working for simplicity to lowest order in $\bm{\beta}=\mathbf{v}/c$, Minkoswki's laboratory-frame constitutive relations for an isotropic dielectric with rest-frame permittivity $\chi'(\omega')$ write~\cite{Minkowski1908}
\begin{subequations}
\begin{gather}
\mathbf{B} = -c^{-2}\mathbf{v}\times\chi'(\omega')\mathbf{E}+{\mu_0}\mathbf{H}
\label{Eq:BMink}\\
\mathbf{D} = \epsilon_0[1+\chi'(\omega')]\cdot\mathbf{E}+c^{-2}\chi'(\omega')\mathbf{v}\times\mathbf{H}.
\label{Eq:DMink}
\end{gather}
\end{subequations}
Simply taking the curl of Eq.~\eqref{Eq:BMink} and using $\bm{\nabla}\times\mathbf{H} = \partial\mathbf{D}/\partial t$ and $\bm{\nabla}\times\mathbf{E} = -\partial \mathbf{B}/\partial t$ and leads to
\begin{multline}
-\mu_0\frac{\partial^2}{\partial t^2}\mathbf{D} = \bm{\nabla}\times\bm{\nabla}\times\mathbf{E}-i\frac{\partial}{\partial t} c^{-2} \bm{\nabla}\times\chi'(\omega')\mathbf{v}\times\mathbf{E}. 
\end{multline}
Now $\mathbf{v}\times\mathbf{H}$ on the right hand side of Eq.~(\ref{Eq:DMink}) can to lowest order in $\mathbf{v}/c$ be approximated by $\mathbf{v}\times\mathbf{B}$. Putting these pieces together we get
\begin{align}
\bm{\nabla}&\times\bm{\nabla}\times\mathbf{E} = -c^{-2}\left[1+\chi'(\omega')\right]\frac{\partial^2}{\partial t^2}\mathbf{E}\nonumber\\
 & +c^{-2}\chi'(\omega')\frac{\partial}{\partial t}\left(\bm{\nabla}\times\mathbf{v}\times\mathbf{E} +\mathbf{v}\times\bm{\nabla}\times\mathbf{E}\right)
\label{Eq:WaveEqE_int}
\end{align}
Considering now the particular case of a rotation at constant angular frequency $\bm{\Omega}$ and a wave propagating along the rotation axis of this medium (i.~e. $\mathbf{k}\parallel\bm{\Omega}$), one further finds that
\begin{equation}
\bm{\nabla}\times\mathbf{v}\times\mathbf{E} +\mathbf{v}\times\bm{\nabla}\times\mathbf{E} = \bm{\Omega}\times\mathbf{E} + \mathcal{O}(\beta^2)
\end{equation}
so that finally
\begin{equation}
c^{2}\bm{\nabla}\times\bm{\nabla}\times\mathbf{E} = -\left[1+\chi'(\omega')\right]\frac{\partial^2}{\partial t^2}\mathbf{E} +\chi'(\omega')\bm{\Omega}\times\frac{\partial}{\partial t}\mathbf{E}
\end{equation}
to first order in $\beta$.

\section{Wave equation for $\mathbf{E}$ in a moving unmagnetized plasma from the laboratory-frame perturbative method}
\label{App:WaveELab}

Consider a moving cold neutral unmagnetized plasma made of ions (charge $q_i=e$) and electrons (charge $q_e=-e$) and assume and unspecified equilibrium such that equilibrium quantities verify $n_{e0}=n_{i0}=n_0=cst$ (i.~e. homogeneous background) and $\mathbf{u}_{e0}= \mathbf{u}_{i0}=\bm \beta c$ (i.~e. electrons and ions flow are locally equal). We study the response of this moving plasma to a small perturbation using a two-fluid model made of Maxwell-Amp{\`e}re and Maxwell-Faraday equations, plus momentum and charge conservation equations for each species $s$. The linearized system of equations for the perturbation $(\mathbf{E},\mathbf{B},n_{s1},\mathbf{u}_{s1})$ then writes
\begin{subequations}\label{eq:system_non-magnetized_plasma}
    \begin{gather}\label{eq:nonmag1}
        \partial_{ct}\mathbf{E}-c\bm\nabla\times\mathbf{B}+\frac{n_0e}{\varepsilon_0 c}\left(\mathbf{u}_{i1}-\mathbf{u}_{e1}\right)+\frac{e}{\varepsilon_0}\left(n_{i1}-n_{e1}\right)\bm\beta=0\\\label{eq:nonmag2}
        \bm\nabla\times\mathbf{E}+\partial_t\mathbf{B}=0\\\label{eq:nonmag3}
        \left(\partial_{ct}+\bm \beta\cdot\bm\nabla+(\bm\nabla\otimes\bm \beta)^T\right) \bm{u}_{s1}-\frac{q_{s}}{m_{s}}\bm{E}/c-\frac{q_{s}}{m_{s}}\bm\beta \times \bm{B}=0\\\label{eq:nonmag4}
        n_{s0}\bm\nabla\cdot\bm u_{s1}/c+\left(\partial_{ct}+(\bm\nabla\cdot\bm \beta)+\bm \beta\cdot\bm\nabla\right)n_{s1}=0
    \end{gather}
\end{subequations}
Using Eq.~\eqref{eq:nonmag2} to eliminate $\mathbf{B}$ and then applying the operator $\partial_{ct}\left(\partial_{ct}+\bm \beta\cdot\bm\nabla+(\bm\nabla\otimes\bm \beta)^T\right)$ to Eq.~\eqref{eq:nonmag1}, we get to first order in $\beta$
\begin{multline}\label{eq:step1}
    \left(\partial_{ct}+\bm \beta\cdot\bm\nabla+(\bm\nabla\otimes\bm \beta)^T\right)\left(\partial_{ct}^2\mathbf{E}+\bm\nabla\times\bm\nabla\times\mathbf{E}\right) \\
    +\frac{en_0}{\varepsilon_0c}\partial_{ct}\left(\partial_{ct}+\bm \beta\cdot\bm\nabla+(\bm\nabla\otimes\bm \beta)^T\right)\left(\mathbf{u}_{i1}-\mathbf{u}_{e1}\right)\\
    +\frac{e}{\varepsilon_0}\partial_{ct}^2\left(n_{i1}-n_{e1}\right)\bm \beta=0.
\end{multline}
Meanwhile, Eq.~\eqref{eq:nonmag4} gives, also to first order in $\beta$, 
\begin{equation}
\frac{e}{\varepsilon_0}\partial_{ct}^2\left(n_{i1}-n_{e1}\right)\bm\beta=-\frac{en_0}{\varepsilon_0c}\bm\nabla\cdot\partial_{ct}(\mathbf{u}_{i1}-\mathbf{u}_{e1})\bm\beta,
\end{equation}
which, using Eq.~\eqref{eq:nonmag3} to zeroth order in $\beta$ i.~e.
\begin{equation}
\partial_{ct}\bm u_{s1}=\frac{q_s}{m_s}\mathbf{E},
\end{equation}
makes it possible to rewrite the last term in Eq.~\eqref{eq:step1} as 
\begin{equation}\label{eq:step2}
        \frac{e}{\varepsilon_0}\partial_{ct}^2\left(n_{i1}-n_{e1}\right)\bm\beta=-\omega_p^2\left(\bm\beta\otimes\bm\nabla\right)\mathbf{E}
\end{equation}
Moreover, using Eqs.~\eqref{eq:nonmag3} and \eqref{eq:nonmag2}, the second to last term in Eq.~\eqref{eq:step1} rewrites
\begin{multline}\label{eq:step3}
        \frac{en_0}{\varepsilon_0c}\partial_{ct}\left(\partial_{ct}+\bm \beta\cdot\bm\nabla+(\bm\nabla\otimes\bm \beta)^T\right)\left(\mathbf{u}_{i1}-\mathbf{u}_{e1}\right)\\=\omega_p^2\left(\partial_{ct}\mathbf{E}/c-\bm\beta\times\bm\nabla\times\mathbf{E}\right)
\end{multline}
Using finally Eqs.~\eqref{eq:step2} and \eqref{eq:step3} in Eq.~\eqref{eq:step1}, and the identity 
\begin{equation}
(\bm\nabla\otimes\bm\beta)^T\mathbf{E}+(\bm\nabla\cdot\mathbf{E})\bm\beta=\bm\nabla\times\bm\beta\times\mathbf{E}+(\bm\nabla\cdot\bm\beta)\mathbf{E}+\bm\beta\cdot\bm\nabla\mathbf{E},
\end{equation}
finally yields
\begin{multline}
\partial^2_{ct}\bm\nabla\times\bm\nabla\times\mathbf{E}+\left(\partial^2_{ct}+\omega_p^2\right)\partial_{ct}^2\mathbf{E}\\
    -\omega_p^2\left(\bm\beta\times\bm\nabla\times\partial_{ct}\mathbf{E}+\bm\nabla\times\bm\beta\times\partial_{ct}\mathbf{E}\right)\\
    -2\omega_p^2\bm\beta\cdot\bm\nabla\partial_{ct}\mathbf{E}-\omega_p^2(\bm\nabla\cdot\bm\beta)\partial_{ct}\mathbf{E}=0.
\end{multline}
Assuming a harmonic time dependence $e^{-i\omega t}$ and dividing this last equation by $-\omega^2$ we finally obtain the wave equation for $\mathbf{E}$ for a moving unmagnetized plasma given in Eq.~\eqref{Eq:WaveEqUnmagnetizedPlasma}.

\end{document}